\documentclass[a4paper,12pt]{article}
\pdfoutput=1
\usepackage{hyperref} 

\usepackage{geometry} 
\usepackage{amsmath}
\usepackage{amssymb}
\usepackage{color}
\usepackage{graphicx}
\usepackage{booktabs}
\usepackage[small,bf]{caption}
\hypersetup{pdfauthor={D. Marzocca, S.T. Petcov, A. Romanino, M. C. Sevilla}, pdftitle={Nonzero Ue3 from Charged Lepton Corrections and the Atmospheric Neutrino Mixing Angle}, colorlinks, bookmarksnumbered, citecolor=darkblue, linkcolor=darkblue, pdfstartview=FitH, urlcolor=darkblue, filecolor=darkblue, linktocpage}

\definecolor{darkblue}{cmyk}{1,0.5,0,0.2}

\newcommand{\arXhref}[1]{\href{http://arxiv.org/abs/arXiv:#1}{#1}} 

\newcommand{\be}{\begin{equation}}
\newcommand{\ee}{\end{equation}}

\newcommand{\eq}[1]{eq.~(\ref{eq:#1})}

\newcommand{\eqs}[1]{eqs.~(\ref{eq:#1})}

\newcommand{\nohyphens}%
       {\hyphenpenalty=10000\exhyphenpenalty=10000\relax}

\DeclareMathOperator{\diag}{diag}

\def\pM{\ensuremath{\genfrac{}{}{0pt}{1}{+}{\scriptstyle(\kern-1pt-\kern-1pt)}}}

\def\gtap{\mathrel{ \rlap{\raise 0.511ex \hbox{$>$}}{\lower 0.511ex
   \hbox{$\sim$}}}} 
\def\ltap{\mathrel{ \rlap{\raise 0.511ex
   \hbox{$<$}}{\lower 0.511ex \hbox{$\sim$}}}}

\allowdisplaybreaks[1]

\newlength{\myem}
\newcounter{mysubequation}[equation]

\newcommand{\SISSA}{SISSA/ISAS and INFN, \\
  Via Bonomea 265, I--34136 Trieste, Italy}

\newcommand{\preprintnumber}{%
SISSA  03/2013/FISI\\
IFIC/13-03}
 
\newcommand{\titletext}
{Nonzero $|U_{e3}|$ from Charged Lepton Corrections
and the Atmospheric Neutrino Mixing Angle}
\newcommand{\authortext}{
\large
David~Marzocca$^{\, a}$,~
S.~T.~Petcov$^{\, a, b}$
\footnote{Also at: 
Institute of Nuclear Research and Nuclear Energy, 
Bulgarian Academy of Sciences, 1784 Sofia, Bulgaria.},
Andrea~Romanino$^{\, a}$,~
M.~C.~Sevilla$^{\, c}$
\medskip\\\em\normalsize 
$\mbox{}^a$ \SISSA
\\[0.1\baselineskip]
$\mbox{}^b$ Kavli IPMU (WPI), University of Tokyo, \\ 5-1-5 Kashiwanoha, 277-8583 Kashiwa, Japan
\\[0.1\baselineskip] 
$\mbox{}^c$ 
Instituto de Fisica Corpuscular, CSIC-Universitat de Valencia, \\
Apartado de Correos 22085, 
E-46071 Valencia, Spain}

\title{
\normalsize
\vspace{-3cm}
\hspace*{\fill}
\begin{tabular}[t]{l}\preprintnumber\end{tabular}
\vspace{3\baselineskip}\\ {\Large\bfseries\titletext}
{\normalsize\bfseries(updated using the results of the global fits of 2013 data\footnote{The Addendum on pages \pageref{Addendum}--\pageref{Addendum_out} is not present in the published version of this paper.} )}
\bigskip}

\author{\begin{minipage}[t]{0.8\textwidth}
\normalsize\centering\authortext
\end{minipage}}
\date{}

\begin{document}


\maketitle


\begin{abstract}
\noindent
After the successful determination of the reactor 
neutrino mixing angle \mbox{$\theta_{13} \cong 0.16 \neq 0$}, 
a new feature suggested by the current 
neutrino oscillation data is a sizeable deviation of the 
atmospheric neutrino mixing angle $\theta_{23}$ from $\pi/4$.
Using the fact that the neutrino mixing matrix 
$U = U^\dagger_{e}U_{\nu}$, where $U_{e}$ and $U_{\nu}$ 
result from the diagonalisation of the charged lepton 
and neutrino mass matrices, and assuming that  $U_{\nu}$ 
has a i) bimaximal (BM), ii) tri-bimaximal (TBM) form, 
or else iii) corresponds to the conservation of the 
lepton charge $L' = L_e - L_\mu - L_{\tau}$ (LC), we 
investigate quantitatively what are the minimal forms 
of $U_e$, in terms of angles and phases it contains, 
that can provide the requisite corrections to 
$U_{\nu}$ so that $\theta_{13}$, $\theta_{23}$
and the solar neutrino mixing angle  $\theta_{12}$ 
have values compatible with the current data.
Two possible orderings of the 12 and the 23
rotations in $U_e$, ``standard'' and ``inverse'', 
are considered. The results we obtain depend strongly 
on the type of  ordering. In the case of ``standard''
ordering, in particular, the Dirac CP violation phase 
$\delta$, present in $U$, is predicted to have a value 
in a narrow interval around i) $\delta \cong \pi$ in the 
BM (or LC) case, ii) $\delta \cong 3\pi/2$ or $\pi/2$ 
in the TBM case, the CP conserving values 
$\delta = 0, \pi, 2\pi$ being excluded in the TBM case 
at more than $4\sigma$. 

\noindent In the addendum we discuss the implications of the latest 2013 data.
\end{abstract}

\newpage



\section{Introduction}

 Understanding the origin of the patterns of neutrino
masses and mixing, emerging from the neutrino oscillation,
$^3H$ $\beta-$decay, etc. data is one of the most
challenging problems in neutrino physics.
It is part of the more general fundamental problem
in particle physics of understanding the origins of
flavour, i.e., of the patterns
of the quark, charged lepton and neutrino masses
and of the quark and lepton mixing.

     At present we have compelling evidence for the
existence of mixing of three light massive neutrinos
$\nu_i$, $i=1,2,3$, in the weak charged lepton current
(see,~e.g.,~\cite{PDG12}). The masses $m_i$ of the three
light neutrinos $\nu_i$ do not exceed approximately
1 eV, $m_i \ltap 1$ eV, i.e., they are much smaller than the
masses of the charged leptons and quarks.
The three light neutrino mixing we will concentrate on 
in the present article, is described (to a good approximation)
by the Pontecorvo, Maki, Nakagawa, Sakata (PMNS) $3\times 3$
unitary mixing matrix, $U_{\rm PMNS}$. In the widely used
standard parametrisation \cite{PDG12},  $U_{\rm PMNS}$ is
expressed in terms of the solar, atmospheric and reactor
neutrino mixing angles $\theta_{12}$,  $\theta_{23}$ and
$\theta_{13}$, respectively, and one Dirac - $\delta$, and
two Majorana \cite{BHP80} - $\alpha_{21}$ and $\alpha_{31}$,
CP violation (CPV) phases:
\be
 U_{\text{PMNS}} \equiv U = V(\theta_{12},\theta_{23},\theta_{13},\delta)\,
Q(\alpha_{21},\alpha_{31})\,,
\label{eq:UPMNS}
\ee
%
where
\be 
V = \left(
     \begin{array}{ccc}
       1 & 0 & 0 \\
       0 & c_{23} & s_{23} \\
       0 & -s_{23} & c_{23} \\
     \end{array}
   \right)\left(
            \begin{array}{ccc}
              c_{13} & 0 & s_{13}e^{-i \delta} \\
              0 & 1 & 0 \\
              -s_{13}e^{i \delta} & 0 & c_{13} \\
            \end{array}
          \right)\left(
                   \begin{array}{ccc}
                     c_{12} & s_{12} & 0 \\
                     -s_{12} & c_{12} & 0 \\
                     0 & 0 & 1 \\
                   \end{array}
                 \right)\,,
\label{eq:V}
\ee
%
\be
Q =  \diag(1, e^{i \alpha_{21}/2}, e^{i \alpha_{31}/2})\,,
\label{Q}
\ee
%
and we have used the standard notation
$c_{ij} \equiv \cos\theta_{ij}$,
$s_{ij} \equiv \sin\theta_{ij}$
with $0\leq \theta_{ij}\leq \pi/2$,
$0\leq \delta \leq 2\pi$ and,
in the case of interest for our analysis,
$0\leq \alpha_{j1}\leq 2\pi$, $j=2,3$
(see, however, \cite{EMSPEJP09}).
If CP invariance holds, we have
$\delta =0,\pi$, and \cite{LW81}
$\alpha_{21(31)} = 0,\pi$.

The neutrino oscillation data, accumulated over many years,
allowed to determine the parameters which drive the solar and
atmospheric neutrino oscillations,
$\Delta m^{2}_{21}$, $\theta_{12}$ and
$|\Delta m^{2}_{31}| \cong |\Delta m^{2}_{32}|$, $\theta_{23}$, 
with a high precision (see, e.g., \cite{Nu2012}).

Furthermore, there were  spectacular developments
in the last 1.5 years in what concerns the angle $\theta_{13}$
(see, e.g., \cite{PDG12}). They culminated
in a high precision determination of $\sin^22\theta_{13}$
in the Daya Bay experiment with reactor
$\bar{\nu}_e$ \cite{An:2012eh}:
\begin{equation}
 \sin^22\theta_{13} = 0.089 \pm 0.010 \pm 0.005\,.
\label{DBayth13}
\end{equation}
%
Similarly the RENO,  Double Chooz, and T2K
experiments reported, respectively, $4.9\sigma$, $2.9\sigma$ and
$3.2\sigma$ evidences for a non-zero value of $\theta_{13}$
\cite{RENODCT2Kth13}, compatible with the Daya Bay result.
The high precision measurement on $\theta_{13}$ 
described above and the fact that $\theta_{13}$ 
turned out to have a relatively large value, 
have far reaching implications
for the program of research in neutrino physics 
(see, e.g., \cite{PDG12}).
After the successful measurement of $\theta_{13}$,
the determination of  the absolute neutrino mass scale, 
of the type of the neutrino mass spectrum, 
of the nature - Dirac or Majorana, of massive neutrinos, 
as well as getting information about 
the status of CP violation in the lepton sector, 
are the most pressing and challenging problems 
and the highest priority goals of the research in the field of 
neutrino physics.

 A global analysis of the latest  
neutrino oscillation data presented at 
the Neutrino 2012 International 
Conference \cite{Nu2012}, was performed 
in \cite{Fogli:2012ua}. The results on $\sin^2\theta_{12}$,  
$\sin^2\theta_{23}$ and $\sin^2\theta_{13}$ 
obtained in \cite{Fogli:2012ua}, which play important 
role in our further discussion, are given 
in Table \ref{tab:globalfit_PMNS_angles}.
\begin{table}
\hspace{-0.5cm}
\begin{tabular}{l | c c c c} 
\toprule
Parameter & Best fit	& $1 \sigma$ range & $2 \sigma$ range & $3 \sigma$ range \\ \hline
$\sin \theta_{13}$ & 0.155 & 0.147 - 0.163 & 0.139 - 0.170 & 0.130 - 0.177 \\
$\sin^2 \theta_{12}$  & 0.307 & 0.291 - 0.325 & 0.275 - 0.342 & 0.259 - 0.359 \\
$\sin^2\theta_{23}$ (NH) & 0.386 &	0.365 - 0.410 & 0.348 - 0.448 & 0.331 - 0.637 \\
$\sin^2\theta_{23}$ (IH) &  0.392 & 0.370 - 0.431 & 0.353 - 0.484 $\oplus$ 0.543 - 0.641& 0.335 - 0.663 \\
$\delta$ (NH) & 3.39 & 2.42 - 4.27 & --- & --- \\
$\delta$ (IH) & 3.42 & 2.61 - 4.62 & --- & --- \\
\bottomrule
\end{tabular}
\caption{Summary of the results of the global 
fit of the PMNS mixing angles taken from 
\cite{Fogli:2012ua} and used in our analysis.
The results on the atmospheric neutrino angle $\theta_{23}$ and on the Dirac CPV phase $\delta$
depend on the type of neutrino mass hierarchy. 
The values of $\sin^2\theta_{23}$ and $\delta$ obtained in 
both the cases of normal hierarchy (NH) 
and inverted hierarchy (IH) are shown. 
\label{tab:globalfit_PMNS_angles}
}
\end{table}
%
An inspection of Table \ref{tab:globalfit_PMNS_angles} 
shows that, in addition to the nonzero value of $\theta_{13}$, 
the new feature which seems to be suggested by the 
current global neutrino oscillation data 
is a sizeable deviation of the angle $\theta_{23}$ 
from the value $\pi/4$. This trend is confirmed 
by the results of the subsequent analysis of the 
global neutrino oscillation 
data performed in \cite{CGGMSchw12}. 

   Although $\theta_{13}\neq 0$, $\theta_{23} \neq \pi/4$ 
and $\theta_{12} \neq \pi/4$, 
the deviations from these values are small, in fact
we have $\sin\theta_{13}\cong 0.16 \ll 1$,
$\pi/4 - \theta_{23} \cong 0.11$ and 
\mbox{$\pi/4 - \theta_{12} \cong 0.20$},
where we have used the relevant best fit values in Table 
\ref{tab:globalfit_PMNS_angles}. 
The value of $\theta_{13}$ and the 
magnitude of deviations of $\theta_{23}$ 
and $\theta_{12}$ from $\pi/4$
suggest that the observed values of 
$\theta_{13}$, $\theta_{23}$ and 
$\theta_{12}$ might originate from 
certain ``symmetry'' values which 
undergo relatively small (perturbative) 
corrections as a result of 
the corresponding symmetry breaking.
This idea was and continues to be 
widely explored in attempts to 
understand the pattern of  
mixing in the lepton sector (see, e.g., 
\cite{GTani02,FPR04,SPWR04,Romanino:2004ww,HPR07,Marzocca:2011,Alta,Chao:2011sp}). 
Given the fact that the PMNS matrix is a product 
of two unitary matrices, 
\begin{equation} 
U = U_e^\dagger\, U_\nu\,,
\label{UeUnu1}
\end{equation}
%
where $U_e$ and $U_\nu$ result respectively from the diagonalisation 
of the charged lepton and neutrino mass matrices, it is usually assumed that 
$U_\nu$ has a specific form dictated by a symmetry which 
fixes the values of the three mixing angles in 
$U_\nu$ that would differ, in general, by perturbative 
corrections from those measured in the 
PMNS matrix, while $U_e$ (and symmetry breaking effects 
that we assume to be subleading) provide the requisite corrections. 
A variety symmetry forms of $U_\nu$ have been explored 
in the literature on the subject (see, e.g., \cite{AlbRode2010}).
In the present study we will consider three widely used forms.\\
i) Tribimaximal Mixing (TBM) \cite{TBM}:
\begin{eqnarray}
\label{TBM1}
U_{\rm TBM} = 
\left(
\begin{array}{ccc}
\sqrt{\frac{2}{3}} & \sqrt{\frac{1}{3}} & 0 \\[0.2cm]
-\sqrt{\frac{1}{6}} & \sqrt{\frac{1}{3}} & \sqrt{\frac{1}{2}}  
\\[0.2cm]
\sqrt{\frac{1}{6}} & -\sqrt{\frac{1}{3}} & \sqrt{\frac{1}{2}}  
\end{array}
\right)\,;
\end{eqnarray}
%
ii) Bimaximal Mixing (BM) \cite{BM}:
\begin{eqnarray}
\label{BM1}
U_{\rm BM} = 
\left(
\begin{array}{ccc}
\frac{1}{\sqrt{2}} &  \frac{1}{\sqrt{2}} 
&  0 \\[0.3cm]
-\frac{1}{2} &  \frac{1}{2} &  \frac{1}{\sqrt{2}} \\[0.3cm]
\frac{1}{2} &  -\frac{1}{2} &  \frac{1}{\sqrt{2}} \\[0.3cm]
\end{array}
\right)\,;
\end{eqnarray}
%
iii) the form of $U_\nu$ resulting from 
the conservation of the lepton charge 
$L' = L_e - L_\mu - L_{\tau}$ of the 
neutrino Majorana mass matrix \cite{SPPD82} (LC):
\begin{eqnarray}
\label{Lprime}
U_{\rm LC} = 
\left(
\begin{array}{ccc}
\frac{1}{\sqrt{2}} &  \frac{1}{\sqrt{2}} 
&  0 \\[0.3cm]
-\frac{c^\nu_{23}}{\sqrt{2}} &  \frac{c^\nu_{23}}{\sqrt{2}} & s^\nu_{23}\\[0.3cm]
\frac{s^\nu_{23}}{\sqrt{2}} & -\,\frac{s^\nu_{23}}{\sqrt{2}} & c^\nu_{23}\\[0.3cm]
\end{array}
\right)\,,
\end{eqnarray}
%
where $c^\nu_{23} = \cos\theta^\nu_{23}$ 
and  $s^\nu_{23} = \sin\theta^\nu_{23}$. 

We will define the assumptions we make on 
$U_e$ and $U_\nu$ in full generality in Section~\ref{sec:Setup}. 
Those assumptions allow us to cover, in particular, the case 
of corrections from $U_e$ to the three widely used forms of $U_\nu$ 
indicated above. We would like to notice here that
if  $U_e = {\bf 1}$, ${\bf 1}$ 
being the unity $3\times 3$ matrix, 
we have:\\
i) $\theta_{13} = 0$ in all three cases 
of interest of $U_\nu$;\\
ii) $\theta_{23}= \pi/4$, 
if $U_\nu$ coincides with 
$U_{\rm TBM}$ or $U_{\rm BM}$, 
while  $\theta_{23}$ can have an 
arbitrary value if $U_\nu$ is given 
by $U_{\rm LC}$;\\
iii) $\theta_{12}= \pi/4$, 
for $U_\nu= U_{\rm BM}$ or $U_{\rm LC}$, 
while $\theta_{12}= \sin^{-1}(1/\sqrt{3})$ 
if $U_\nu= U_{\rm TBM}$.\\ 
Thus, the matrix $U_e$ has to generate 
corrections \\
i) leading to $\theta_{13} \neq 0$ compatible 
with the observations in all three cases of 
$U_\nu$  considered;\\
ii) leading to the observed deviation of 
$\theta_{12}$ from $\pi/4$ in the cases of 
$U_\nu= U_{\rm BM}$ or $U_{\rm LC}$.\\
iii) leading to the sizeable
deviation of $\theta_{23}$ from $\pi/4$
for $U_\nu= U_{\rm TBM}$ or $U_{\rm BM}$,
if it is confirmed by further data 
that $\sin^2\theta_{23} \cong 0.40$.

  In the present article we investigate 
quantitatively what are the ``minimal'' 
forms of the matrix $U_e$ in terms of the 
number of angles and phases it contains, 
that can provide the requisite corrections  
to $U_{\rm TBM}$, $U_{\rm BM}$ and $U_{\rm LC}$
so that the angles in the resulting 
PMNS matrix have values which are 
compatible with those derived from the 
current global neutrino oscillation data, 
Table  \ref{tab:globalfit_PMNS_angles}.
Our work is a natural continuation of the 
study some of us have done in \cite{Marzocca:2011} and earlier 
in \cite{FPR04,SPWR04,Romanino:2004ww,HPR07}.

 The paper is organised as follows.  
In Section~\ref{sec:Setup} we describe the general 
setup and we introduce the two types of ``minimal''
charged lepton ``rotation'' matrix $U_e$ we will consider: 
with ``standard'' and ``inverse'' ordering. 
The two differ by the order in which the 12 and 23 rotations 
appear in $U_e$. In the same Section we derive 
analytic expressions for the mixing angles and the Dirac 
phase $\delta$ of the PMNS matrix in terms of the 
parameters of the charged lepton matrix $U_e$ both 
for the tri-bimaximal and bimaximal (or LC) forms of 
the neutrino ``rotation'' matrix $U_{\nu}$. 
In Sections~\ref{sec:Results} and \ref{sec:inverse} 
we perform a numerical analysis 
and derive, in particular, the intervals 
of allowed values at a given C.L. 
of the neutrino mixing angle parameters 
$\sin^2\theta_{12}$, $\sin^2\theta_{23}$ and
$\sin^2\theta_{13}$, the Dirac phase $\delta$ and 
the rephasing invariant $J_{CP}$ associated with $\delta$,
in the cases of the standard and inverse ordering of 
the charged lepton corrections. 
A summary and conclusions are presented in Section~\ref{sec:Conclusions}. 
Further details are reported in two appendices. 
In Appendix~\ref{app:Upmns_param} we illustrate in detail
the parametrisation we use for the standard ordering setup. 
Finally, in Appendix~\ref{app:Stat} we describe the 
statistical analysis used to obtain the numerical results.

%
\section{General Setup}
\label{sec:Setup}
%
%

While neutrino masses and mixings may or may not 
look anarchical, the hierarchy of charged lepton masses 
suggests an ordered origin of lepton flavour. 
Given the wide spectrum of specific theoretical models, 
which essentially allows to account for any pattern 
of lepton masses and mixings, we would like to consider 
here the consequence for lepton mixing of simple, 
general assumptions on its origin. 
As we have indicated in the Introduction,
we are interested in the possibility that the $\theta_{13}$ 
mixing angle originates because of the contribution of 
the charged lepton sector to lepton mixing~
\cite{GTani02,FPR04,SPWR04,Romanino:2004ww,HPR07,Marzocca:2011,Alta}. 
The latter assumption needs a precise definition. 
In order to give it, let us recall
that the PMNS mixing matrix is given by 
\be
\label{eq:U_def}
U = U_e^\dagger U_\nu,\quad \text{with $U_e$, $U_\nu$ defined by} \quad 
\begin{aligned}
m_E = U^*_{e^c} m^\text{diag}_E U^\dagger_e \\
m_\nu = U^*_\nu m^\text{diag}_\nu U^\dagger_\nu 
\end{aligned} ,
\ee
%
where $m_E$ and $m_\nu$ are respectively the charged 
lepton and neutrino Majorana mass matrices 
(in a basis assumed to be defined by the 
unknown physics accounting 
for their structure) and $m^\text{diag}_E$ and 
$m^\text{diag}_\nu$ are diagonal with positive eigenvalues. 

  We will assume that the neutrino contribution 
$U_\nu$ to the PMNS matrix $U$ has $U^\nu_{13} = 0$, 
so that the PMNS angle $\theta_{13}$ vanishes in 
the limit in which the charged lepton 
contribution $U_e$ can be neglected, $U_e = \mathbf{1}$. 
This is a prediction of a number of theoretical models. 
As a consequence, $U_\nu$ can be parameterized as
\begin{equation}
\label{eq:Unu}
U_\nu = \Psi_\nu R_{23}(\theta_{23}^\nu) R_{12}(\theta_{12}^\nu) \Phi_\nu ,
\end{equation}
%
where $R_{ij}(\theta)$ is a rotation by an 
angle $\theta$ in the $ij$ block and 
$\Psi_\nu$, $\Phi_\nu$ are diagonal matrices 
of phases. We will in particular consider specific 
values of $\theta^\nu_{12}$ and, in certain cases, of 
$\theta^\nu_{23}$, representing the predictions 
of well known models.

The above assumption on the structure of $U_\nu$ 
is not enough to draw conclusions on lepton mixing: 
any form of $U$ can still be obtained by combining 
$U_\nu$ with an appropriate charged lepton contribution 
$U_e = U_\nu U^\dagger$. However, the hierarchical 
structure of the charged lepton mass matrix allows 
to motivate a form of $U_e$ similar to that of 
$U_\nu$, with $U^e_{13} = 0$, so that we can write:
\footnote{The use of the inverse in 
eqs. (\ref{eq:Ue}) and (\ref{eq:UeI}) 
is only a matter of convention. This choice allows us to 
lighten the notation in the subsequent expressions.}
\begin{equation}
\label{eq:Ue}
U_e = \Psi_e R^{-1}_{23}(\theta_{23}^e) R^{-1}_{12}(\theta_{12}^e) \Phi_e .
\end{equation}
%

In fact, the diagonalisation of the charged lepton 
mass matrix gives rise to a value of $U^e_{13}$
that is small enough to be negligible for our purposes, 
unless the hierarchy of masses is a consequence of 
correlations among the entries of the charged lepton 
mass matrix or the value of the  element $(m_E)_{31}$, 
contrary to the common lore, happens to be sizable. 
In such a scheme, with no 13 rotation neither in the neutrino 
nor in the charged lepton sector, 
the PMNS angle $\theta_{13}$ is 
generated purely by the interplay of the 23 and 
12 rotations in eqs. (\ref{eq:Unu}) and (\ref{eq:Ue}).

While the assumption that $U^e_{13}$ is small, leading to \eq{Ue}, is well motivated, textures leading to a sizeable $U^e_{13}$ are not excluded. In such cases, it is possible to obtain an ``inverse ordering'' of the $R_{12}$ and $R_{23}$ rotations in $U_e$: 
%
\begin{equation}
\label{eq:UeI}
	U_e = \Psi_e R^{-1}_{12}(\theta_{12}^e) R^{-1}_{23}(\theta_{23}^e) \Phi_e .
\end{equation}
%
In the following, we will  also consider such a possibility.

%
\subsection{Standard Ordering}
\label{sec:StanOrd}
%

Consider first 
the standard ordering in \eq{Ue}.
We can then combine $U_\nu$ and $U_e$ 
in eqs.~(\ref{eq:Unu}) and (\ref{eq:Ue})
to obtain the PMNS matrix. When doing that, 
the two 23 rotations, by the $\theta^\nu_{23}$ and 
$\theta^e_{23}$ angles, can be combined into a 
single 23 rotation by an angle $\hat\theta_{23}$. 
The latter angle is not necessarily simply given by the sum
\mbox{$\hat\theta^{\phantom{e}}_{23} = \theta^\nu_{23} + \theta^e_{23}$}
because of the possible effect of the phases in 
$\Psi_\nu$, $\Psi_e$ 
(see further, eq.~\eqref{eq:theta23hat_relation}).
Nevertheless, the combination $R_{23}(\theta_{23}^e) 
\Psi^*_e \Psi_\nu R_{23}(\theta_{23}^\nu)$ 
entering the PMNS matrix is surely a unitary matrix 
acting on the 23 block and, as such, 
it can be written as $\Omega_\nu R_{23} (\hat\theta_{23}) \Omega_e$, 
where $\Omega_{\nu,e}$ are diagonal matrices of phases and \mbox{$\hat{\theta}_{23} \in [0, \pi/2]$}. 
Moreover, we can write $\Omega_\nu R_{23} (\hat\theta_{23}) \Omega_e =
\Omega'_\nu \Phi R_{23} (\hat\theta_{23}) \Omega'_e$, 
where $\Phi = \diag(1, e^{i \phi}, 1)$ and $\Omega'_{\nu,e}$ 
are diagonal matrices of phases that commute with 
the 12 transformations and either are unphysical 
or can be reabsorbed in other phases. 
The PMNS matrix can therefore be written as 
\cite{Marzocca:2011}
\be
U = P R_{12}(\theta^e_{12}) \Phi 
R_{23}(\hat{\theta}_{23}) R_{12}(\theta_{12}^\nu) Q,
\label{eq:Upmns_param}
\ee
%
where the angle $\hat{\theta}_{23}$ 
can have any value, 
$P$ is a diagonal matrix of unphysical phases, 
$Q$ contains the two Majorana CPV phases, 
and $\Phi = \diag(1, e^{i \phi}, 1)$ 
contains the only Dirac CPV 
phase. The explicit relation between the physical 
parameters $\hat\theta_{23}$, $\phi$ and 
the original parameters of the model 
($\theta^\nu_{23}$, $\theta^e_{23}$, and 
the two phases 
in $\Psi =\Psi^*_e\Psi_\nu$)
can be useful to connect our results to 
the predictions of specific theoretical models. 
We  provide it in Appendix~\ref{app:Upmns_param}. 

The observable angles in the standard PMNS parametrisation are given by
\be \begin{split}
	\sin \theta_{13} &= \left| U_{e3} \right| =  \sin \theta^e_{12} \sin \hat{\theta}_{23}, \\
	\sin^2 \theta_{23} &=  \frac{\left| U_{\mu3} \right|^2}{1- \left| U_{e 3} \right|^2 } = 
					\sin^2 \hat{\theta}_{23} \frac{\cos^2 \theta_{12}^e}{1 - \sin^2 \theta^e_{12} \sin^2 \hat{\theta}_{23} }, \\
	\sin^2 \theta_{12} &= \frac{\left| U_{e2} \right|^2}{1- \left| U_{e3} \right|^2 } = 
					\frac{\left| \sin \theta_{12}^\nu \cos \theta_{12}^e + e^{i \phi} \cos \theta_{12}^\nu \cos \hat{\theta}_{23} \sin \theta_{12}^e \right|^2}{1 - \sin^2 \theta^e_{12} \sin^2 \hat{\theta}_{23} }.
	\label{eq:chlep_corrections}
\end{split} 
\ee
%
The rephasing invariant related to the Dirac CPV phase, 
which determines the magnitude of CP violation effects 
in neutrino oscillations \cite{PKSP3nu88}, 
has the following well known form 
in the standard parametrisation:
\be
J_{CP} = \text{Im} \left\{ U_{e1}^* U_{\mu3}^* U_{e3} U_{\mu1} \right\} = \frac{1}{8} \sin \delta \sin 2 \theta_{13} \sin 2 \theta_{23} \sin 2 \theta_{12} \cos \theta_{13}\,.
\label{eq:Jcp_std}
\ee
%
At the same time, in the parametrisation given in  
\eq{Upmns_param}, we get:
\be
	J_{CP} = -\frac{1}{8} \sin \phi \sin 2 \theta_{12}^e  \sin 2 \hat{\theta}_{23} \sin \hat{\theta}_{23} \sin 2 \theta_{12}^\nu\,.
		\label{eq:Jcp_chlep}
\ee
The relation between the phases $\phi$ and $\delta$ 
present in the two parametrisations is obtained 
by equating \eq{Jcp_std} and \eq{Jcp_chlep} 
and  taking also into account the 
corresponding formulae for the real part 
of $U_{e1}^* U_{\mu3}^* U_{e3} U_{\mu1}$. 
To leading order in $\sin \theta_{13}$,
one finds the approximate relation $\delta \simeq - \phi$ 
(see further eqs. (\ref{sindNOpi4}), 
(\ref{cosdNOpi4}) and eqs. (\ref{sindNOsqrt3}) 
and (\ref{cosdNOsqrt3}) for the exact relations).

\begin{figure}
 \centering
 \includegraphics[width=9cm]{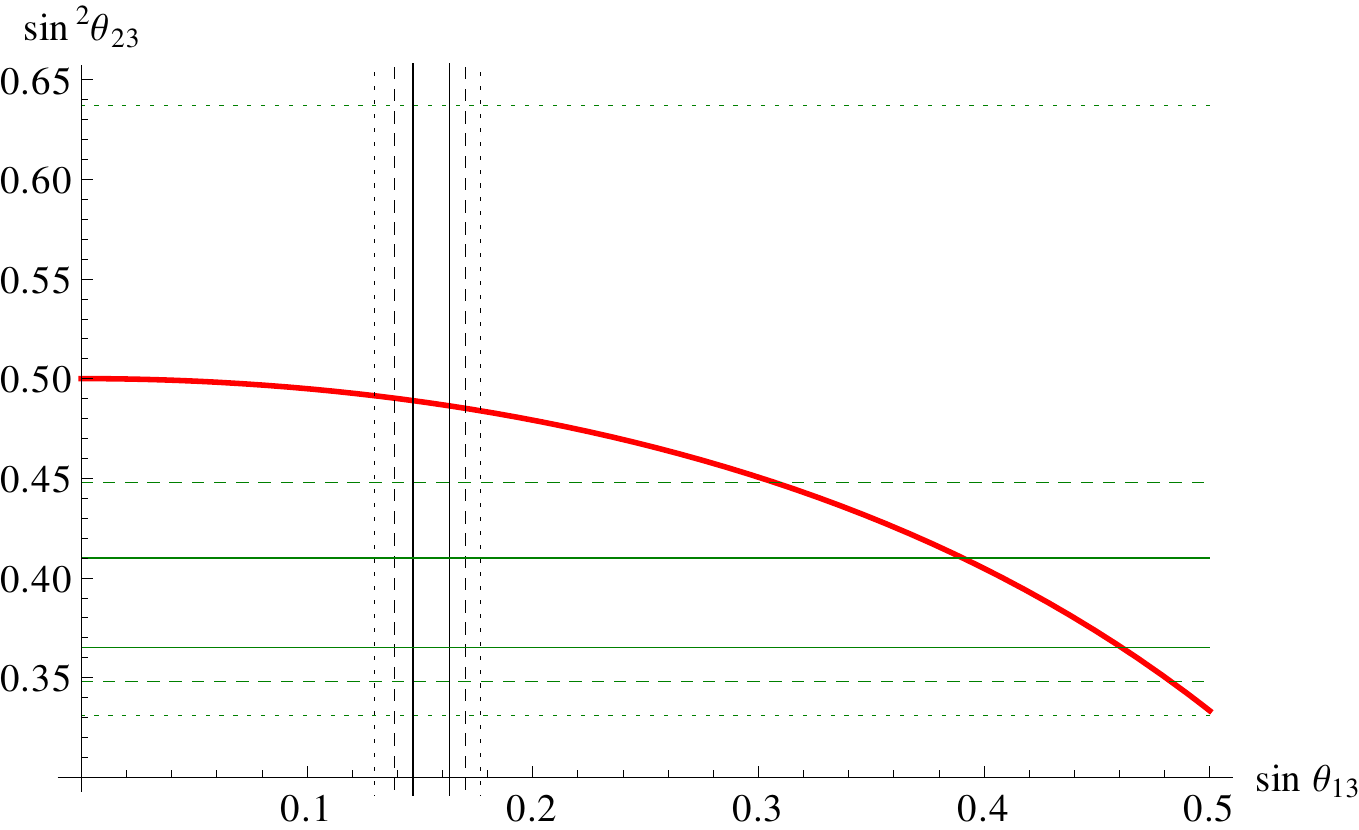}
 \caption{The thick red line corresponds to the relation 
in eq.\eqref{eq:TBM_s13_s23sq}. The black and green lines show the 
$1\sigma, 2\sigma, 3\sigma$ contours 
(solid, dashed and dotted lines, respectively) for 
$\sin \theta_{13}$ and $\sin^2 \theta_{23}$, 
as obtained in \cite{Fogli:2012ua} 
(see Table \ref{tab:globalfit_PMNS_angles}).}
\label{fig:TBM_s13_s23sq}
\end{figure}

  In this work we aim to go beyond the simplest 
cases considered already, e.g., in \cite{Marzocca:2011}, 
where the charged lepton corrections to neutrino 
mixing are dominated only by the angle $\theta_{12}^e$ 
and $\hat\theta_{23}$ is fixed at 
the maximal value $\hat\theta_{23} = \pi/4$, 
and consider the case in which 
$\hat\theta_{23}$ is essentially free. 
A deviation of $\hat\theta_{23}$ 
from $\pi/4$ can occur in models in which 
$\theta_{23}^\nu = \pi/4$ (BM, TBM) because of the charged 
lepton contribution to $\hat\theta_{23}$, or in models in which 
$\theta_{23}^\nu$ itself is not maximal (LC). 
This choice allows to account for a sizeable
deviation of $\theta_{23}$ from the value $\pi/4$,
which appears to be suggested by the data~\cite{Fogli:2012ua}.
If we keep the assumption $\hat\theta_{23} = \pi/4$, 
the atmospheric mixing angle would be given by 
%
\be
\sin^2 \theta_{23} = 
\frac{1}{2}\, \frac{1 - 2 \sin^2 \theta_{13}}{1 - \sin^2 \theta_{13} } 
\cong \frac{1}{2}\,(1 - \sin^2\theta_{13})\,, \quad 
\text{where} \quad \sin \theta_{13} = \frac{1}{\sqrt{2}} \sin \theta_{12}^e\,.
\label{eq:TBM_s13_s23sq}
\ee
%
This in turn would imply that the deviation from 
maximal atmospheric neutrino mixing corresponding 
to the observed value of $\theta_{13}$ is relatively small, 
as shown in Fig. \ref{fig:TBM_s13_s23sq}.
%
%
As for the neutrino angle $\theta^\nu_{12}$, we will consider two cases:
%
\begin{itemize}
	\item \emph{bimaximal} mixing (BM): 
$\theta^\nu_{12} = \displaystyle \frac{\pi}{4}$ 
(as also predicted by models with approximate conservation of 
$L^\prime = L_e - L_\mu - L_\tau$ );
 \item \emph{tri-bimaximal} mixing (TBM): 
$\theta^\nu_{12} = \displaystyle\sin^{-1}\frac{1}{\sqrt{3}}$.
\end{itemize}
%
Since in the approach we are following the four parameters 
of the PMNS matrix (the three measured angles $\theta_{12}$, 
 $\theta_{23}$, $\theta_{13}$ and the CPV Dirac phase $\delta$) 
will be expressed in terms of only three parameters 
(the two angles $\theta^e_{12}$, $\hat\theta_{23}$ and the phase $\phi$),
the values of $\theta_{12}$,  $\theta_{23}$, $\theta_{13}$ and 
$\delta$ will be correlated. More specifically, $\delta$
can be expressed as a function of the three angles, 
$\delta = \delta(\theta_{12},\theta_{23},\theta_{13})$,
and its value will be determined by the values of the angles.
As a consequence, the $J_{CP}$ factor also will be a function 
of $\theta_{12}$, $\theta_{23}$ and $\theta_{13}$, which will allow us 
to obtain predictions for the magnitude of the CP violation effects 
in neutrino oscillations using the current data on 
$\sin^2\theta_{12}$, $\sin^2\theta_{23}$ and $\sin\theta_{13}$.

 We note first that using eq. (\ref{eq:chlep_corrections}) 
we can express $\sin^2\theta_{23}$ in terms of 
$\sin^2\hat\theta_{23}$ and $\sin^2\theta_{13}$:
%
\be
\sin^2 \theta_{23} = 
\frac{\sin^2\hat\theta_{23} - \sin^2 \theta_{13}} 
{1 - \sin^2\theta_{13}}\,,~~\cos^2 \theta_{23} = 
\frac{\cos^2\hat\theta_{23}} 
{1 - \sin^2\theta_{13}}\,.
\label{eq:s23sqNO}
\ee
%
It follows from these equations that $\hat\theta_{23}$ differs little from 
$\theta_{23}$ (it is somewhat larger).
Further, using eqs. (\ref{eq:chlep_corrections}) and 
(\ref{eq:s23sqNO}), we can express  $\sin^2\theta_{12}$
in terms of $\theta^{\nu}_{12}$, $\theta_{23}$, $\theta_{13}$ and $\phi$:
%
\begin{eqnarray}
\nonumber	
\sin^2\theta_{12} &= 
\left (1 - \cos^2\theta_{23}\cos^2\theta_{13}\right )^{-1}\,
\left [ \sin^2\theta^{\nu}_{12}\sin^2\theta_{23} + 
\cos^2\theta^{\nu}_{12}\cos^2\theta_{23}\sin^2\theta_{13} \right.  \\[0.3cm]
&+ \left. \frac{1}{2}\, 
\sin2\theta^{\nu}_{12}\sin2\theta_{23}\sin\theta_{13}\cos\phi \right] 
\,.
\label{eq:s12sqNO}
\end{eqnarray}
%
As we have already indicated, we will use in the 
analysis which follows two specific values of 
$\theta^{\nu}_{12} = \pi/4$ (BM or LC); $\sin^{-1}(1/\sqrt{3})$ (TBM). 
Equation (\ref{eq:s12sqNO}) will lead in each of the two cases 
to a new type of ``sum rules'', i.e., to a correlation between 
the value of $\theta_{12}$ and the values  
of $\theta_{23}$, $\theta_{13}$ and $\phi$.
In the case of bimaximal and tri-bimaximal $\theta^{\nu}_{12}$, 
the sum rules have the form:
%
\begin{align}
\label{eq:s12sqNOBM1}
{\rm BM: } \quad&
\sin^2\theta_{12}= \frac{1}{2}\, + \frac{1}{2}\, 
\frac{\sin2\theta_{23}\sin\theta_{13}\cos\phi }
{1 - \cos^2\theta_{23}\cos^2\theta_{13}}
\\
\label{eq:s12sqNOBM2}
&\cong \frac{1}{2} + \cot\theta_{23}\sin\theta_{13}\cos\phi
\left (1 - \cot^2\theta_{23}\sin^2\theta_{13} + 
\mathcal{O}(\cot^4\theta_{23}\sin^4\theta_{13})\right )\,,\\
\label{eq:s12sqNOTBM1}
{\rm TBM: } \quad&
\sin^2\theta_{12} = 
\frac{1}{3}\left (2  + 
\frac{ \sqrt{2}\,\sin2\theta_{23}\sin\theta_{13}\cos\phi - \sin^2\theta_{23}}
{ 1 - \cos^2\theta_{23}\cos^2\theta_{13} } \right )
\\
& \cong \frac{1}{3}\,
\left [ 1 + 2\sqrt{2}\,\cot\theta_{23}\sin\theta_{13}\cos\phi\,
\left (1 - \cot^2\theta_{23}\sin^2\theta_{13}\right) \right.
\nonumber \\
\label{eq:s12sqNOTBM2}
& + \left. \cot^2\theta_{23}\sin^2\theta_{13} + 
\mathcal{O}(\cot^4\theta_{23}\sin^4\theta_{13})\right ]\,.
\end{align}
%
The expressions for $\sin^2\theta_{12}$ 
in eqs. (\ref{eq:s12sqNOBM1}) and 
(\ref{eq:s12sqNOTBM1}) are exact, while those 
given in eqs. (\ref{eq:s12sqNOBM2}) and (\ref{eq:s12sqNOTBM2}) 
are obtained as expansions in the small parameter 
$\cot^2\theta_{23}\sin^2\theta_{13}$.
The latter satisfies $\cot^2\theta_{23}\sin^2\theta_{13} \lesssim 0.063$ 
if  $\sin^2\theta_{23}$ and $\sin^2\theta_{13}$ 
are varied in the  $3\sigma$ intervals 
quoted in Table \ref{tab:globalfit_PMNS_angles}. 
To leading order in $\sin\theta_{13}$ the sum rule in 
eq. (\ref{eq:s12sqNOBM2}) was derived in \cite{SPWR04}.

We note next  that since $\theta_{12}$, 
$\theta_{23}$ and $\theta_{13}$
are known, eq. (\ref{eq:s12sqNO}) allows us 
to express $\cos\phi$ as a function of 
$\theta_{12}$, $\theta_{23}$ and $\theta_{13}$
and to obtain the range of possible values of 
$\phi$. Indeed, it follows from eqs. 
(\ref{eq:s12sqNOBM1}) and (\ref{eq:s12sqNOTBM1}) that
%
\begin{align}
\label{cosphiNOpi4}
{\rm BM: } \qquad& \cos\phi = -\,\frac{\cos2\theta_{12}\,(1 - \cos^2\theta_{23}\cos^2\theta_{13})}
{\sin2\theta_{23}\,\sin\theta_{13}}\,,&\\[0.30cm]
\label{cosphiNOsqrt3}
{\rm TBM: }  \qquad& \cos\phi = \frac{(3\sin^2\theta_{12}-2)\,
(1 - \cos^2\theta_{23}\cos^2\theta_{13}) + \sin^2\theta_{23}}
{\sqrt{2}\sin2\theta_{23}\,\sin\theta_{13}}\,.
\end{align}
%
Taking for simplicity for the best fit values of 
the three angles in the PMNS matrix 
$\sin^2\theta_{12} = 0.31$, 
$\sin^2\theta_{23} = 0.39$ and $\sin\theta_{13} = 0.16$ 
(see Table \ref{tab:globalfit_PMNS_angles}), we get:
%
\begin{equation}
\cos\phi \cong -\,0.99\,~{\rm (BM)};\qquad
\cos\phi \cong -\,0.20\,,~{\rm (TBM)}.
\label{cosphibf}
\end{equation}

Equating the imaginary and real parts of 
$U_{e1}^* U_{\mu 3}^* U_{e3} U_{\mu 1}$ 
in the standard parametrisation and in the parametrisation 
under discussion one can obtain a relation between 
the CPV phases $\delta$ and $\phi$.
We find for the BM case ($\theta^{\nu}_{12}=\pi/4$):
%
\begin{align}
\label{sindNOpi4}
\sin\delta =&\; -\, \frac{\sin\phi}{\sin2\theta_{12}}\,,
\\[0.30cm]
\label{cosdNOpi4}
\cos\delta =&\; \frac{\cos\phi}{\sin2\theta_{12}}\,
\left ( \frac{2\,\sin^2\theta_{23}}
{\sin^2\theta_{23}\cos^2\theta_{13} + \sin^2\theta_{13}} - 1 \right )\,.
\end{align}
%
Since, as can be easily shown,
\begin{equation}
\sin2\theta_{12} = 
\left ( 1 - 4 \frac{\cot^2\theta_{23}\sin^2\theta_{13}\cos^2\phi}
{(1 + \cot^2\theta_{23}\sin^2\theta_{13})^{2}} \right)^{\frac{1}{2}}\,,
\label{eq:sin2th12pi4}
\end{equation}
%
we indeed have to leading order in $\sin\theta_{13}$, 
$\sin\delta \cong -\sin\phi$ and $\cos\delta \cong \cos\phi$.

The expressions for $\sin\delta$ and $\cos\delta$ in 
eqs. (\ref{sindNOpi4}) and (\ref{cosdNOpi4}) are exact. 
It is not difficult to check that
we have  $\sin^2\delta + \cos^2\delta = 1$.
Using the result for $\cos\phi$, eq. (\ref{cosphiNOpi4}), 
we can get expressions for $\sin\delta$ and $\cos\delta$ 
in terms of $\theta_{12}$, $\theta_{23}$ and $\theta_{13}$.
We give below the result for $\cos\delta$:
\begin{equation}
\cos\delta = -\, \frac{1}{2\sin\theta_{13}}\,
\cot2\theta_{12}\,\tan\theta_{23}\,
\left ( 1 - \cot^2\theta_{23}\,\sin^2\theta_{13}\right )\,.
\label{2cosdNOpi4}
\end{equation}
%
Numerically we find for $\sin^2\theta_{12} = 0.31$, 
$\sin^2\theta_{23} = 0.39$ and $\sin\theta_{13} = 0.16$:
\begin{equation}
\sin\delta \cong \pm 0.170\,,~~\cos\delta \cong -\,0.985\,.
\end{equation}
%
Therefore, we have $\delta \simeq \pi$. 
For fixed $\sin^2\theta_{12}$ and  $\sin\theta_{13}$, 
$|\cos\delta|$ increases with the increasing of 
$\sin^2\theta_{23}$.
However, $\sin^2\theta_{23}$ cannot increase arbitrarily 
since eq.~(\ref{eq:s12sqNOBM1}) and the measured values 
of  $\sin^2\theta_{12}$ and  $\sin^2\theta_{13}$ 
imply that the scheme with bimaximal mixing 
under discussion can be self-consistent only 
for values of $\sin^2\theta_{23}$, which 
do not exceed a certain maximal value. 
The latter is determined  
taking into account the uncertainties in the values of 
$\sin^2\theta_{12}$ and  $\sin\theta_{13}$
in Section 3, where we perform a 
statistical analysis using the data  
on $\sin^2\theta_{23}$, $\sin^2\theta_{12}$,  
$\sin\theta_{13}$ and $\delta$ as given in 
\cite{Fogli:2012ua}.

In a similar way we obtain for the TBM case  
($\theta^{\nu}_{12}=\sin^{-1}(1/\sqrt{3})$):
\begin{align}
\label{sindNOsqrt3}
\sin\delta =&\; -\, \frac{2\sqrt{2}}{3}\,\frac{\sin\phi}{\sin2\theta_{12}}\,, 
\\[0.30cm]
\cos\delta =&\; \frac{2\sqrt{2}}{3\sin2\theta_{12}}\,\cos\phi\,
\left (-1 + \frac{2\sin^2\theta_{23}}
{\sin^2\theta_{23}\cos^2\theta_{13} + \sin^2\theta_{13}}\,\right ) 
\nonumber \\[0.30cm] 
 &\;+ \frac{1}{3\sin2\theta_{12}}\, 
\frac{\sin2\theta_{23}\, \sin\theta_{13}}
{\sin^2\theta_{23}\cos^2\theta_{13} + \sin^2\theta_{13}}\,.
\label{cosdNOsqrt3}
\end{align}
%
The results for $\sin\delta$ and $\cos\delta$   
we have derived are again exact and, 
as can be shown, satisfy 
$\sin^2\delta + \cos^2\delta = 1$.
Using the above expressions and the expression for $\sin^2\theta_{12}$
given in eq. (\ref{eq:s12sqNOTBM1}) and  neglecting the corrections 
due to $\sin\theta_{13}$, we obtain  
$\sin\delta \simeq -\,\sin \phi$ and $\cos\delta \simeq \cos\phi$.
With the help of eq. (\ref{cosphiNOsqrt3}) we can express 
$\sin\delta$ and $\cos\delta$ in terms of $\theta_{12}$, $\theta_{23}$ 
and $\theta_{13}$. The result for $\cos\delta$ reads:
%
\begin{equation}
\cos\delta =  \frac{\tan\theta_{23}}{3\sin2\theta_{12}\sin\theta_{13}}\,
\left [ 1 + \left (3\sin^2\theta_{12} - 2 \right )\, 
 \left (1 - \cot^2\theta_{23}\,\sin^2\theta_{13}\right )\right ]\,.
\label{2cosdNOsqrt3}
\end{equation}
%
For the best fit values of $\sin^2\theta_{12} = 0.31$, 
$\sin^2\theta_{23} = 0.39$ and $\sin\theta_{13} = 0.16$,  we find:
%
\begin{equation}
\sin\delta \cong \pm 0.999\,,~~\cos\delta \cong -\,0.0490\,.
\end{equation}
%
Thus, in this case $\delta \simeq \pi/2$ or $3\pi/2$.
For $\sin^2\theta_{23} = 0.50$ and the same values of 
$\sin^2\theta_{12}$ and $\sin^2\theta_{13}$ we get 
$\cos\delta \cong -0.096$ and $\sin\delta \cong \pm 0.995$.

  The fact that the value of the Dirac CPV phase 
$\delta$ is determined (up to an ambiguity of 
the sign of $\sin\delta$) by the values of the 
three mixing angles  $\theta_{12}$, $\theta_{23}$ 
and $\theta_{13}$ of the PMNS matrix, 
eqs. (\ref{2cosdNOpi4}) and  (\ref{2cosdNOsqrt3}),
are the most striking predictions of the 
scheme considered with standard ordering 
and bimaximal and tri-bimaximal mixing 
in the neutrino sector. 
For the best fit values of $\theta_{12}$, 
$\theta_{23}$ and $\theta_{13}$ we get 
$\delta \cong \pi$ and $\delta \cong \pi/2$ or $3\pi/2$
in the cases of bimaximal and tri-bimaximal mixing, 
respectively. 
These results imply also that in the scheme with 
standard ordering under discussion,  
the $J_{CP}$ factor which determines the magnitude 
of CP violation in neutrino oscillations 
is also a function of the three angles 
$\theta_{12}$, $\theta_{23}$ and $\theta_{13}$ 
of the PMNS matrix:  
\begin{equation} 
J_{CP} = J_{CP}(\theta_{12},\theta_{23},\theta_{13}, 
\delta(\theta_{12},\theta_{23},\theta_{13})) = 
J_{CP}(\theta_{12},\theta_{23},\theta_{13})\,. 
\label{JCPNO}
\end{equation}
%
This allows to obtain predictions for the range of 
possible values of $J_{CP}$ using the current data on 
$\sin^2\theta_{12}$, $\sin^2\theta_{23}$ 
and $\sin\theta_{13}$. 
We present these predictions in Section 3. 
The predictions we derive for $\delta$ and $J_{CP}$ 
will be tested in the experiments searching 
for CP violation in neutrino oscillations, 
which will provide information on the value of 
the Dirac phase $\delta$.

We would like to note that the sum rules we obtain in the BM (LC) and TBM cases, eqs. \eqref{2cosdNOpi4} and \eqref{2cosdNOsqrt3}, differ from the sum rules derived in \cite{Hernandez:2012ra} using van Dyck and Klein type discrete symmetries ($S_4$, $A_4$, $A_5$, etc.), and in \cite{King:2013eh} on the basis of $SU(5)$ GUT and $S_4$, $A_4$ and $\Delta(96)$ symmetries. More specifically, for the values of $\sin^2\theta_{12}$, $\sin^2\theta_{23}$ and $\sin^2\theta_{13}$, compatible with current global neutrino oscillation data, for instance, the predictions for the value of the CPV phase $\delta$ obtained in the present study differ from those found in \cite{Hernandez:2012ra,King:2013eh}. The same comment is valid also for the possible ranges of values of $\sin^2\theta_{12}$ and $\sin^2\theta_{23}$ found by us and in \cite{Hernandez:2012ra}. Our predictions for $\delta$ agree with the ones reviewed in \cite{King:2013eh} in the context of charged lepton corrections, once we take the particular case $\hat \theta_{23} = \theta_{23}^\nu$.


%
\subsection{Inverse Ordering}
\label{sec:InvOrd}
%
%
As anticipated, we also study for completeness 
the case where the diagonalisation of the charged 
lepton mass matrix gives rise to the inverse ordering in \eq{UeI}. 
The PMNS matrix, in this case, can be written 
as \cite{FPR04}
\be
\label{eq:UIO}
U = R_{23}( \tilde \theta^e_{23}) R_{12}(\tilde \theta^e_{12}) 
\Psi R_{23}(\theta^\nu_{23}) R_{12}(\theta^\nu_{12}) \tilde Q,
\ee
%
where unphysical phases have been eliminated, 
$\tilde Q$ contains the two Majorana phases, 
and $\Psi = \diag(1, e^{i \psi}, e^{i \omega})$. 
Unlike in the case of standard ordering, it is not possible to 
combine the 23 rotation in the neutrino and charged lepton 
sector and describe them with a single parameter, $\hat\theta_{23}$. 
After fixing $\theta^\nu_{23}$ and $\theta^\nu_{12}$, we therefore have, 
in addition to the Majorana phases, four independent physical parameters, 
two angles and two phases, one more with respect to 
the case of standard ordering. 
In particular, it is not possible anymore to write 
the mixing matrix in terms of one physical Dirac CPV phase only. 
Thus, in this case the four parameters of the PMNS matrix  
(the three angles $\theta_{12}$, $\theta_{23}$ and 
$\theta_{13}$ and the Dirac CPV phase $\delta$) will 
be expressed in terms of the four parameters of the inverse 
ordering parametrisation of the PMNS matrix, eq. (\ref{eq:UIO}).
We have for $\sin \theta_{13}$, $\sin \theta_{23}$ and 
$\sin \theta_{12}$: 
\be 
\label{eq:chlep_corrections_inv_ord} 
\begin{split}
	\sin \theta_{13} =&\;  \tilde s^e_{12} s^\nu_{23}, \\
	\sin \theta_{23} =&\;  s^\nu_{23} \frac{ \left| (t^\nu_{23})^{-1} \tilde s^e_{23} + e^{i (\psi-\omega)} \tilde c^e_{12} \tilde c^e_{23} \right|}{\sqrt{ 1 - (\tilde s^e_{12} s^\nu_{23} )^2} } , \\
	\sin \theta_{12} =&\;  s^\nu_{12} \frac{\left| \tilde c^e_{12}  +e^{i\psi} (t^\nu_{12})^{-1} \tilde s^e_{12} c^\nu_{23} \right|}{\sqrt{1- (\tilde s^e_{12} s^\nu_{23} )^2 } }. 
\end{split}
\ee
%
Given that the expressions for $\theta_{23}$ and 
$\theta_{13}$ do not depend on the value of $\theta_{12}^\nu$, 
they will be the same for bimaximal and tri-bimaximal 
mixing (in both cases $\theta^\nu_{23} = \frac{\pi}{4}$):
\begin{align}
\label{eq:chlep_corrections_inv_ord_BMandTBM} 
\sin \theta_{13} &=  \frac{\sin \tilde \theta^e_{12}}{\sqrt{2}}, \\[.2cm]
\label{eq:s23sqIO1}
\sin^2 \theta_{23} &=  \frac{1}{2} \frac{1+\sin 2 \tilde \theta_{23}^e \sqrt{\cos 2 \theta_{13}} \cos \omega^\prime - 2 \sin^2 \theta_{13} \cos^2 \tilde \theta_{23}^e}{\cos^2 \theta_{13}} \\[.2cm]
\label{eq:s23sqIO2}
&\cong \frac{1}{2}\left ( 1 + \sin 2 \tilde \theta_{23}^e\, \cos \omega^\prime 
-\, \cos 2 \tilde \theta_{23}^e\,\sin^2\theta_{13} + \mathcal{O}(\sin^4\theta_{13})\,
\right )\,,
\end{align}
%
where the phase $\omega^\prime = \psi - \omega$.
The expression (\ref{eq:s23sqIO2}) for 
$\sin^2 \theta_{23}$ is approximate, the corrections 
being of the order of  $\sin^4\theta_{13}$ or smaller.

For each value of the phase $\psi$,
any value of $\theta_{13}$ and $\theta_{23}$ in the 
experimentally allowed range at a given C.L.,
can be reproduced for an appropriate choice of 
$\omega^\prime$,
$\theta_{12}^e$ and $\theta_{23}^e$. 
This is not always the case for the solar neutrino 
mixing angle $\theta_{12}$, 
as we will see in Section~\ref{sec:inverse}. 
Using \eqs{chlep_corrections_inv_ord_BMandTBM}, 
$\sin^2\theta_{12}$ can be expressed in terms of 
$\theta_{13}$ and $\psi$ as follows:
\begin{itemize}
	\item \emph{bimaximal} mixing (BM$_\text{IO}$), 
$\theta^\nu_{12} =\displaystyle \frac{\pi}{4}$: 
\begin{align}
\label{eq:s12sqIOBM1}
\sin^2 \theta_{12} &=  \frac{1}{2\cos^2 \theta_{13}} 
\left (1+2 \sin \theta_{13} \sqrt{\cos 2 \theta_{13}} \cos \psi - 
\sin^2 \theta_{13} \right )\\[0.20cm]
\label{eq:s12sqIOBM2}
&\simeq \frac{1}{2} + \sin\theta_{13}\,\cos\psi + 
               \mathcal{O}(\sin^5\theta_{13})\,;
\end{align}
	\item \emph{tri-bimaximal} mixing (TBM$_\text{IO}$), 
$\theta^\nu_{12} =\sin^{-1} \frac{1}{\sqrt{3}}$: 
\begin{align}	
\label{eq:s12sqIOTBM1}
\sin^2 \theta_{12} &=  
\frac{1}{3\cos^2 \theta_{13}} \left (1 + 2 \sqrt{2} \sin \theta_{13} 
\sqrt{\cos 2 \theta_{13}} \cos \psi \right ) \\[0.20cm]
\label{eq:s12sqIOTBM2}
&\simeq \frac{1}{3}\,(1 + \sin^2\theta_{13}) + 
\frac{2\sqrt{2}}{3}\sin\theta_{13}\cos\psi + \mathcal{O}(\sin^4\theta_{13})\,.
\end{align}
\end{itemize}
%
The expressions for $\sin^2\theta_{12}$ in eqs. (\ref{eq:s12sqIOBM1}) and 
(\ref{eq:s12sqIOTBM1}) are exact, while those given in (\ref{eq:s12sqIOBM2}) 
and (\ref{eq:s12sqIOTBM2}) are obtained as expansions in $\sin^2\theta_{13}$
in which the terms up to 
$\mathcal{O}( \sin^4\theta_{13})$ and 
$\mathcal{O}( \sin^3\theta_{13})$, respectively,
were kept. Note that the corrections to the approximate expressions for 
$\sin^2\theta_{12}$ are negligibly small, 
being $\mathcal{O}(\sin^4\theta_{13})$.
This together with eq. (\ref{eq:s12sqIOBM2}) and the 
$3\sigma$ ranges of allowed values of 
$\sin^2\theta_{12}$ and $\sin\theta_{13}$ 
quoted in Table \ref{tab:globalfit_PMNS_angles} suggests that 
the bimaximal mixing scheme considered by us can be compatible 
with the current ($3\sigma$) data on $\sin^2\theta_{12}$ 
and $\sin\theta_{13}$ only for a very limited interval 
of negative values of $\cos\psi$ close to ($-1$).

It follows from eqs.  (\ref{eq:s12sqIOBM1}) and 
(\ref{eq:s12sqIOTBM1}) that the value of  
$\cos\psi$ is determined by the values of the PMNS angles 
$\theta_{12}$ and $\theta_{13}$. At the same time, 
$\sin^2\theta_{23}$ depends on two parameters: 
$\omega^\prime$ and $\theta_{23}^e$. 
This implies that the values of $\omega^\prime$ and $\theta_{23}^e$ 
are correlated, but cannot be fixed individually 
using the data on $\sin^2\theta_{23}$.

It is not difficult to derive also the expressions for the 
$J_{CP}$ factor in terms of the inverse ordering parameters 
in the two cases of values of $\theta^\nu_{12}$ 
of interest:

\begin{align}
\!\!\!{\rm BM: } \;\;\; J_{CP} \simeq& -\,\frac{\sin \theta_{13}}{4} 
\left( \sin \psi \cos 2 \tilde \theta_{23}^e + 
\sin \omega^\prime \cos \psi \sin 2 \tilde \theta_{23}^e \right) + 
\mathcal{O}(\sin^2 \theta_{13})\,,
\label{eq:chlep_corrections_inv_ord_BM} 
\\[0.30cm]
\!\!\!{\rm TBM: } \;\;\; J_{CP} \simeq& 
-\, \frac{\sin \theta_{13}}{3 \sqrt{2}} 
\left( \sin \psi \cos 2 \tilde \theta_{23}^e + 
\sin \omega^\prime \cos \psi \sin 2\tilde \theta_{23}^e \right) + 
\mathcal{O}(\sin^2 \theta_{13})\,.
\label{eq:chlep_corrections_inv_ord_TBM} 
\end{align}
%
 
We have not discussed here the LC case
(conservation of the lepton charge 
$L' = L_e - L_\mu - L_{\tau}$)
as it involves five parameters 
($\theta^e_{23}$, $\theta^e_{12}$, $\theta^\nu_{23}$, 
and two CPV phases). 
At the same time, the ``minimal'' LC case with $\theta^e_{23} = 0$ 
is equivalent to the standard ordering case with BM mixing 
(i.e., with $\theta^\nu_{12} = \pi/4$) analised in 
detail in the previous subsection.

  As in the case of the standard ordering, to obtain the CPV 
phase $\delta$ of the standard parametrisation 
of the PMNS matrix from the variables of these models, 
that is the function 
$\delta = \delta(\psi, \omega, \tilde \theta_{23}^e, \theta_{13})$, 
we equate the imaginary and real parts of 
$U_{e1}^* U_{\mu 3}^* U_{e3} U_{\mu 1}$ 
in the two parametrisations.

%
\section{Results with Standard Ordering}
\label{sec:Results}
%

In the numerical analysis presented here, we use the data 
on the neutrino mixing parameters obtained in the global fit of 
\cite{Fogli:2012ua} to constrain the mixing parameters of 
the setup described in Section~\ref{sec:Setup}. Our goal is first 
of all to derive the allowed ranges for the Dirac phase 
$\delta$, the $J_{CP}$ factor and the atmospheric neutrino 
mixing angle parameter $\sin^2\theta_{23}$. 
We will also obtain the allowed values of $\sin^2\theta_{12}$ and 
$\sin^2\theta_{13}$. We start in this Section by considering the 
standard ordering setup, and in particular the two different 
choices for the angle $\theta_{12}^\nu$:  $\theta_{12}^\nu = \pi/4$ (BM and LC), $\theta_{12}^\nu = \sin^{-1} (1/\sqrt{3})$ (TBM).
\begin{figure}[t]
\begin{center}
\vspace*{-1.5cm}
\fbox{\footnotesize Standard Ordering - Normal Hierarchy} \\[0.5cm]
\hspace*{-0.65cm} 
\begin{minipage}{0.5\linewidth}
\begin{center}
	\hspace*{1cm} 
	\fbox{\footnotesize Tribimaximal} \\[0.05cm]
	\includegraphics[width=75mm]{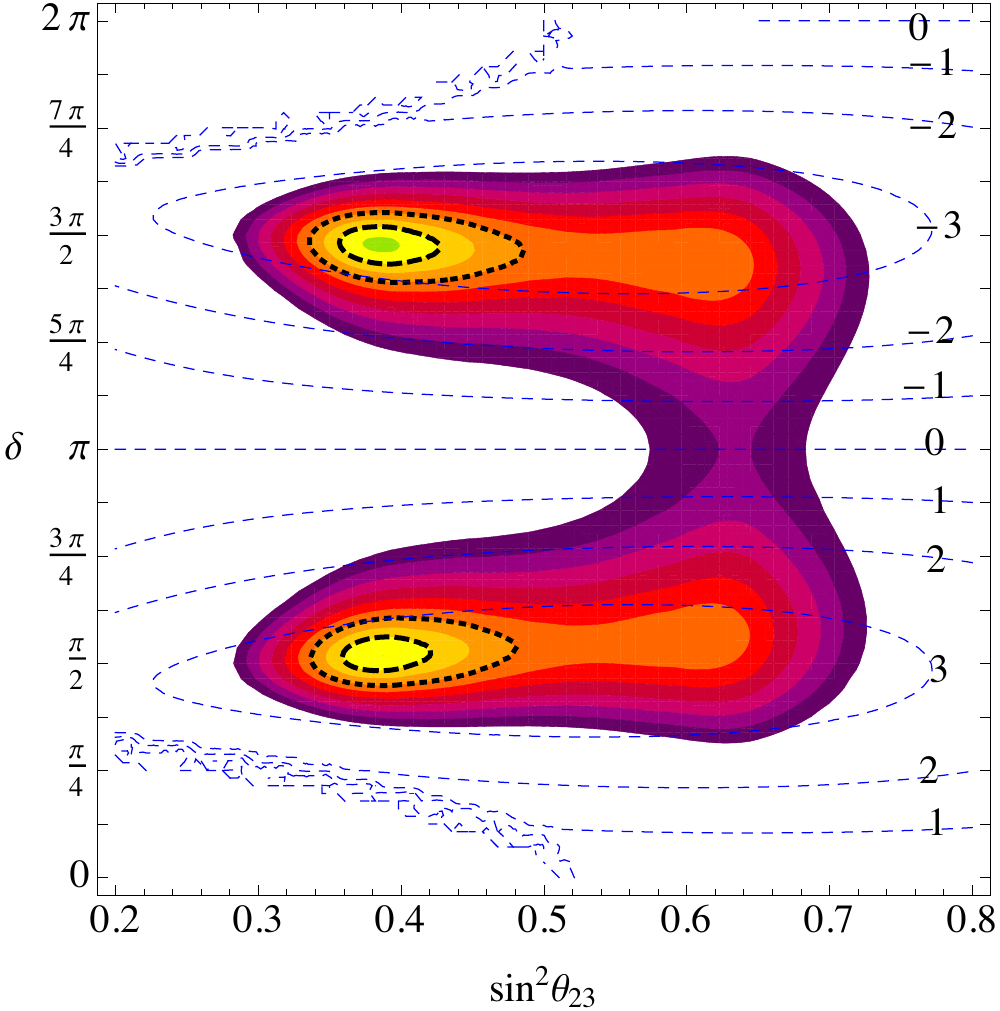}\\
	\vspace*{-0.2cm}
	\mbox{\footnotesize (a)} \\
\end{center}
\end{minipage}
\hspace{0.25cm}
\begin{minipage}{0.5\linewidth}
\begin{center}
	\hspace*{1cm}
	\fbox{\footnotesize Bimaximal} \\[0.05cm]
	\includegraphics[width=75mm]{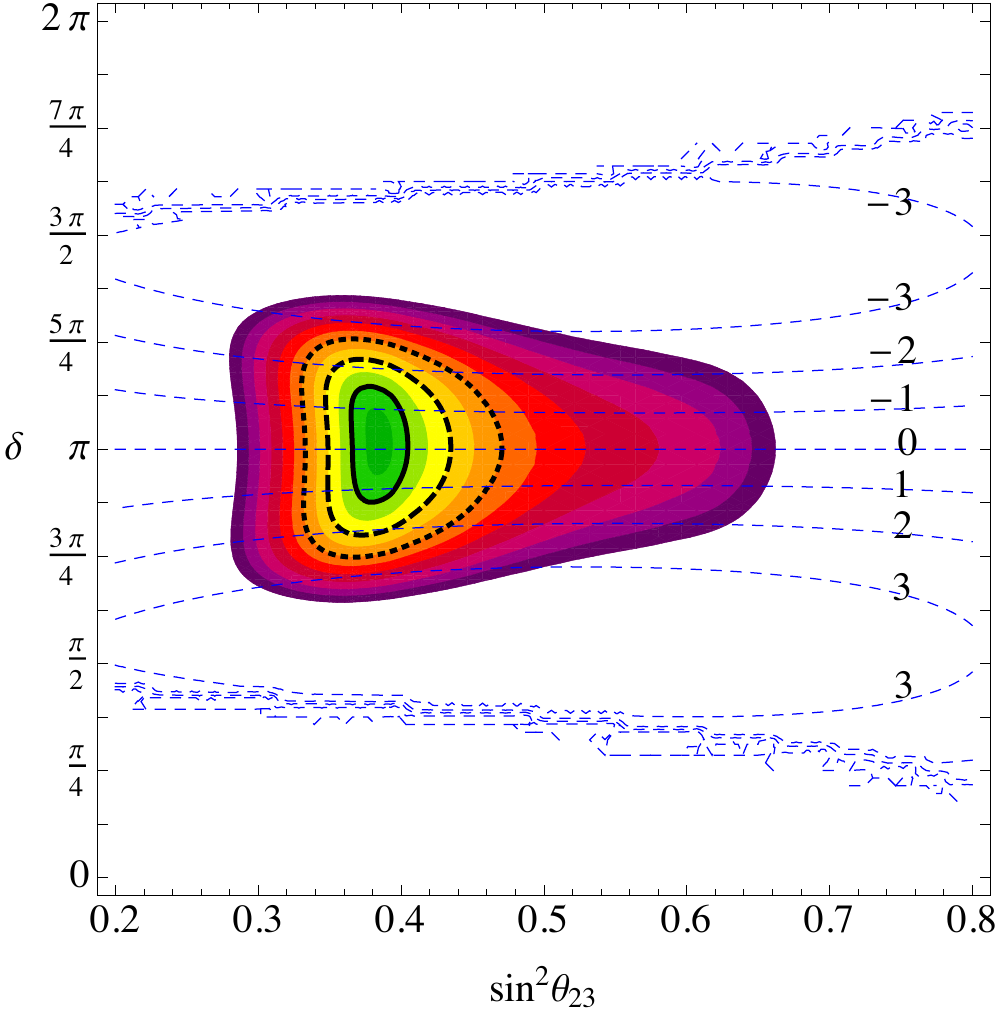}\\
	\vspace*{-0.2cm}
	\mbox{\footnotesize (b)} \\
\end{center}
\end{minipage} 
\\
\begin{minipage}{0.5\linewidth}
\begin{center}
	\includegraphics[width=75mm]{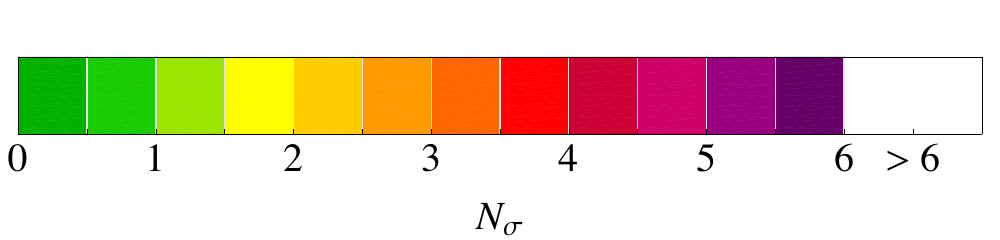}\\
\end{center}
\end{minipage}
\end{center}
\vspace*{-0.5cm}
\caption{\label{fig:results_delta} 
\small
Contour plots for $N_\sigma = \sqrt{\chi^2}$
in the standard ordering setup and normal hierarchy of neutrino masses.
The value of the reactor angle $\theta_{13}$ has been marginalized. 
The solid, dashed and dotted thick lines represent respectively the 
$1\sigma, 2\sigma$ and $3\sigma$ contours. The dashed blue 
lines are contours of constant $\left| J_{CP} \right|$ 
in units of $10^{-2}$.}
\end{figure}
%

 We construct the likelihood function and the $\chi^2$ 
for both schemes of bimaximal and tri-bimaximal 
mixing as described in Appendix \ref{app:Stat}, 
using as parameters for this model $\sin \theta_{13}$, 
$\sin^2 \theta_{23}$ and $\delta$, and exploiting the 
constraints on $\sin^2 \theta_{12}$, $\sin^2 \theta_{23}$, 
$\sin^2 \theta_{13}$ and on $\delta$ 
obtained in \cite{Fogli:2012ua}.

\begin{figure}[p]
 \centering
 \vspace*{-1cm}
 \fbox{\footnotesize Standard Ordering - Tribimaximal} \\[0.5cm]
 \vspace*{-0.2cm} 
 \includegraphics[height=6.5cm]{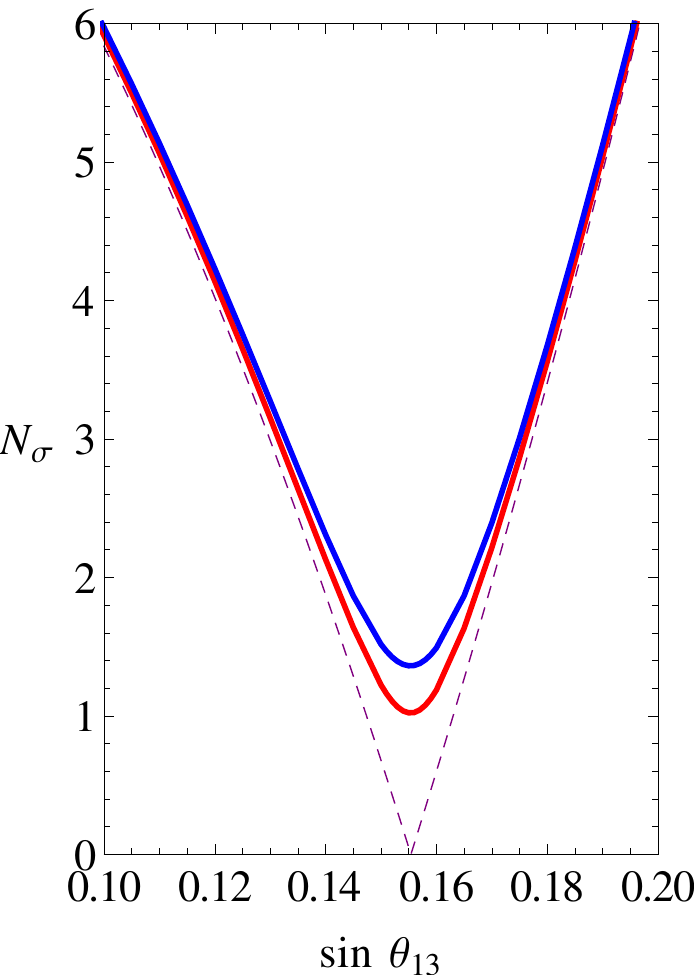} 
 \includegraphics[height=6.5cm]{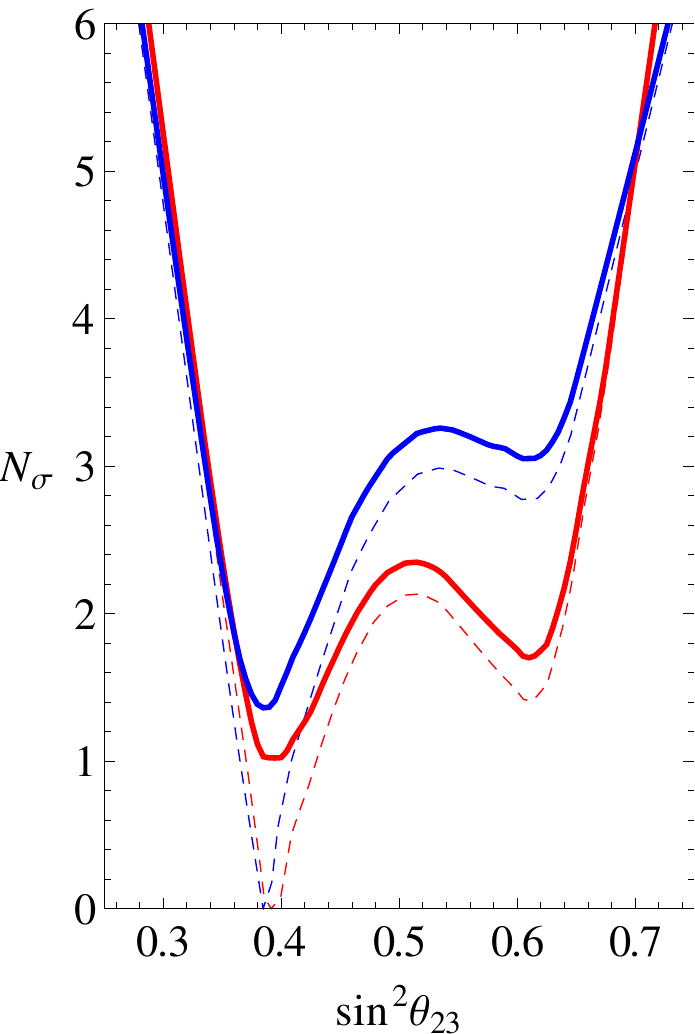}
 \includegraphics[height=6.5cm]{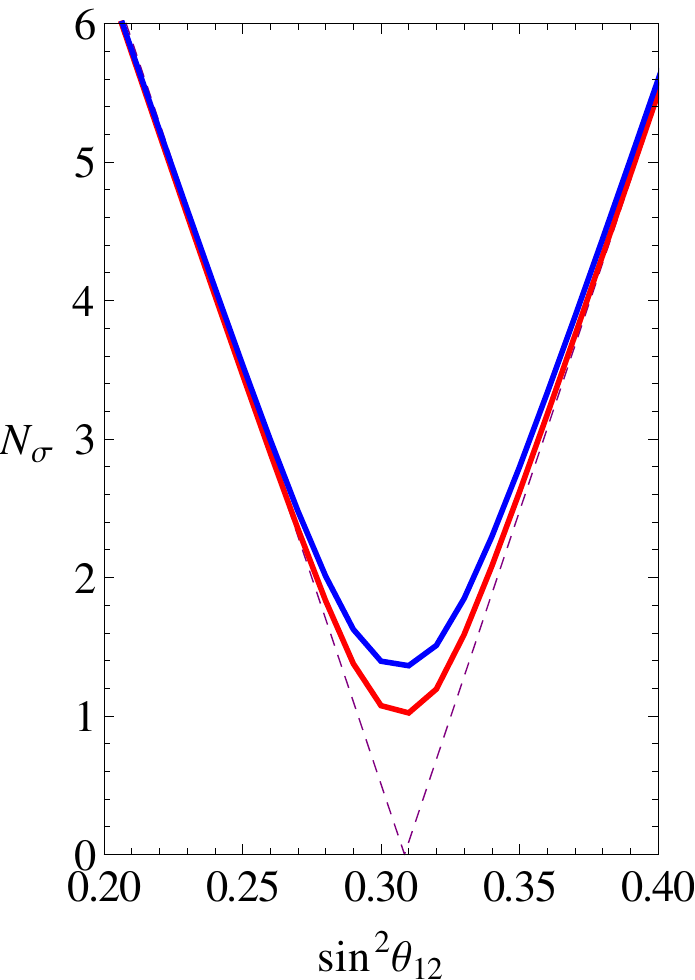}
 \vspace*{0.2cm}
 \fbox{\footnotesize Standard Ordering - Bimaximal} \\[0.5cm]
 \vspace*{-0.2cm} 
 \includegraphics[height=6.5cm]{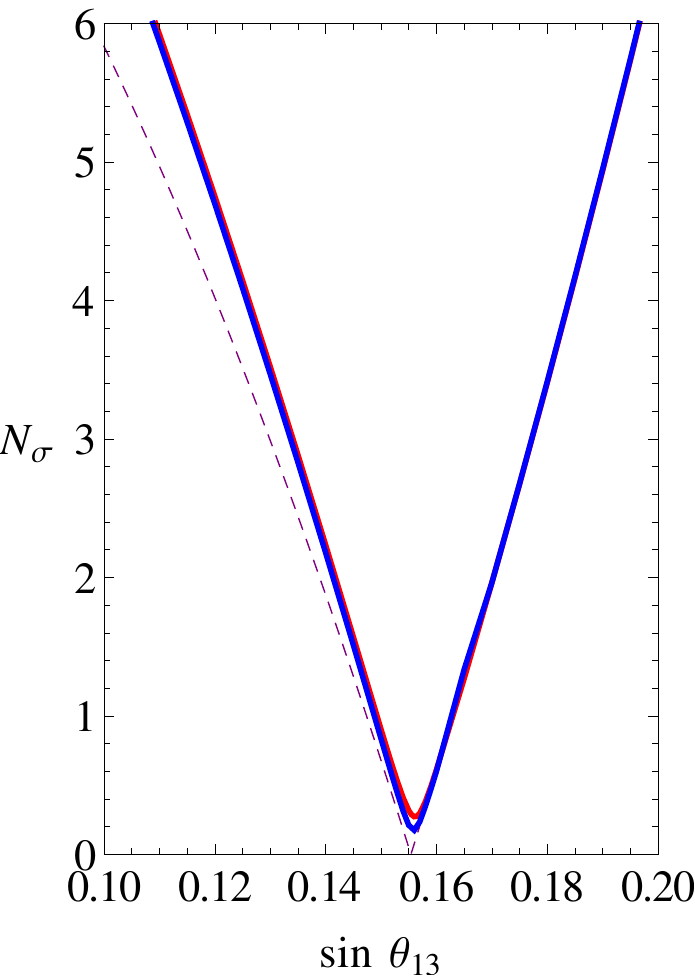} 
 \includegraphics[height=6.5cm]{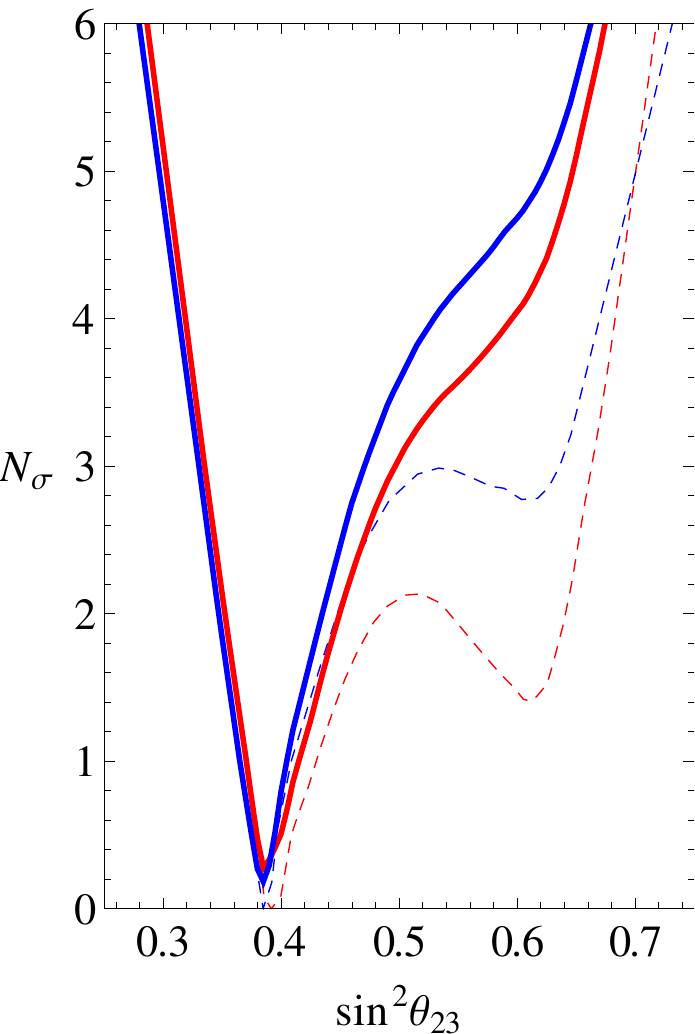}
 \includegraphics[height=6.5cm]{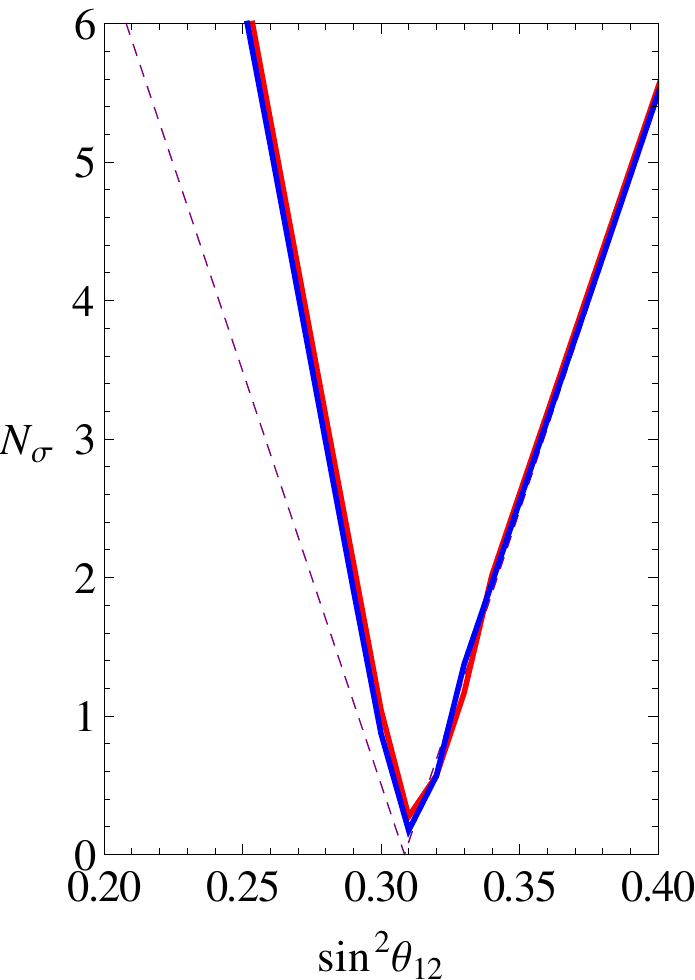}
 \caption{$N_\sigma$ as a function of each mixing angle for 
the TBM and BM models in the standard ordering setup.
The dashed lines represent the results of the global fit 
reported in \cite{Fogli:2012ua} while the thick ones represent 
the results we obtain in our setup. Blue lines are 
for normal hierarchy while the red 
ones are for inverted hierarchy (we used purple when the two bounds are 
approximately identical). 
These bounds are obtained minimizing the value of $N_\sigma$ in 
the parameter space for fixed value of the showed mixing angle.}
\label{fig:Nsigma_bounds_StOrd}
\end{figure}

\begin{figure}[p]
 \centering
 \vspace*{-1cm}
 \fbox{\footnotesize Standard Ordering - Tribimaximal} \\[0.5cm]
 \vspace*{-0.2cm} 
 \includegraphics[height=7cm]{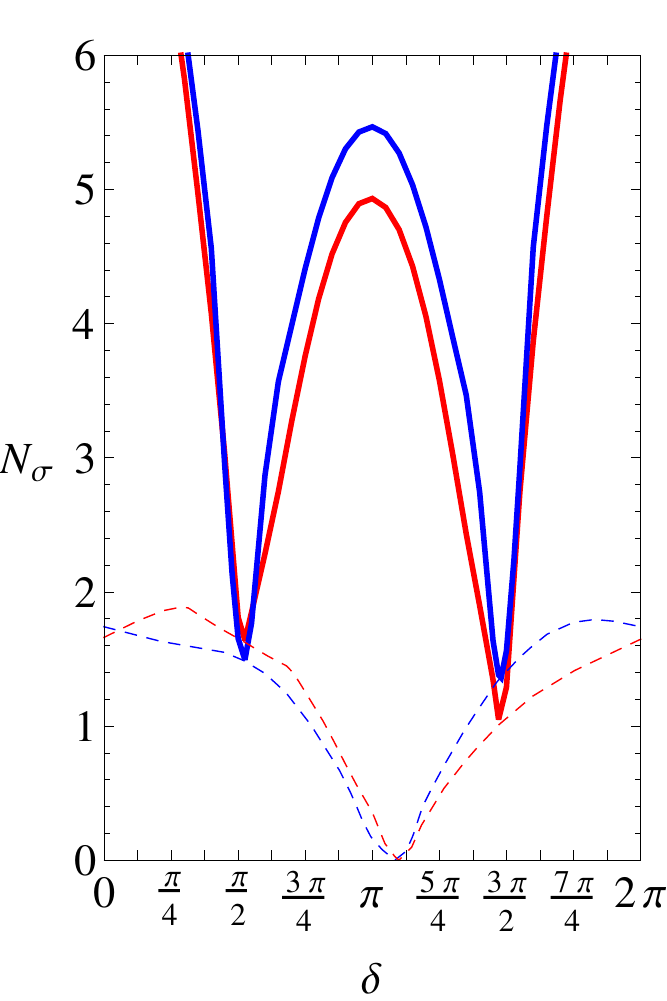} \hspace{1cm}
 \includegraphics[height=6.5cm]{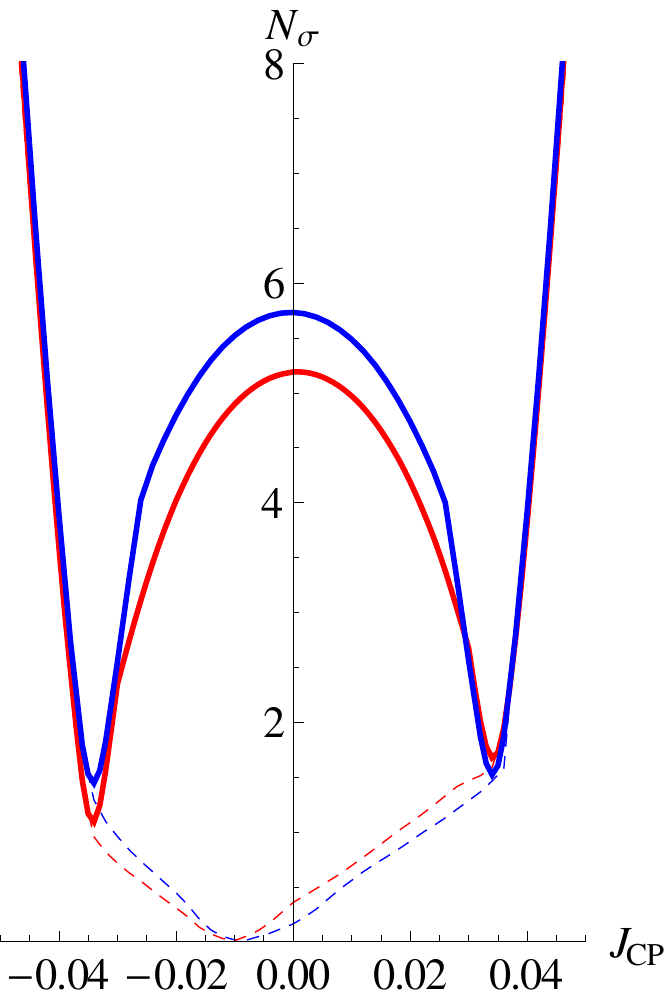}\\
 \vspace*{0.2cm}
 \fbox{\footnotesize Standard Ordering - Bimaximal} \\[0.5cm]
 \vspace*{-0.2cm} 
 \includegraphics[height=7cm]{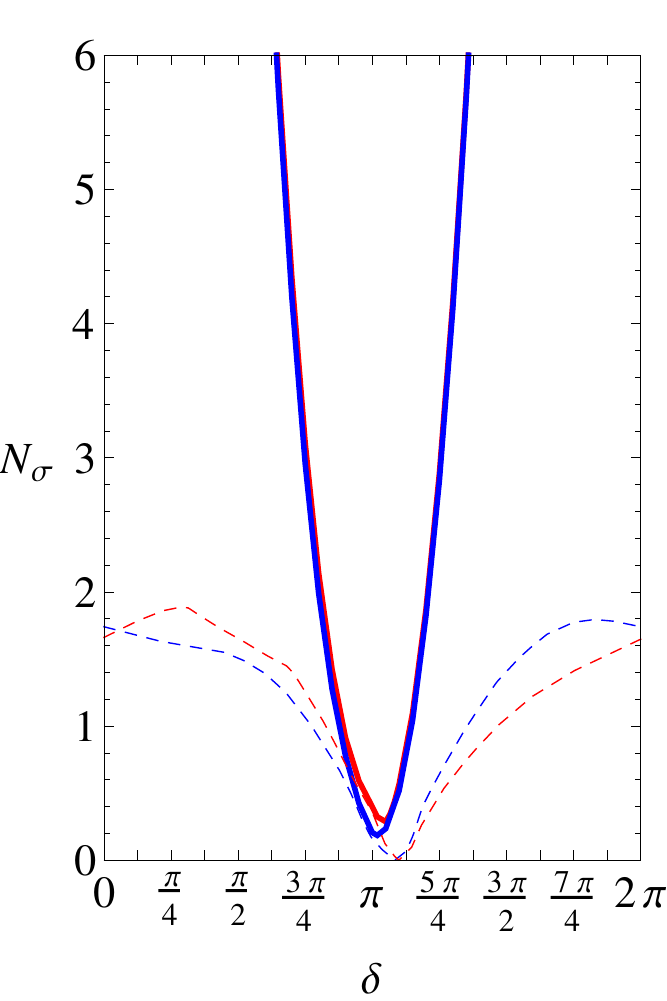} \hspace{1cm}
 \includegraphics[height=6.5cm]{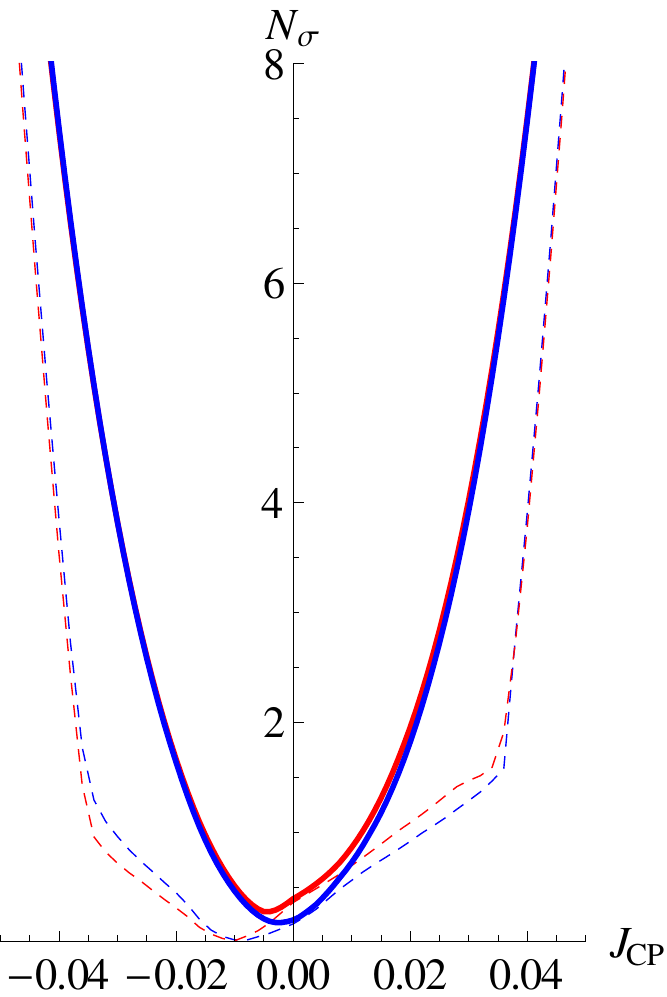}\\
\caption{\label{fig:bounds_Jcp} 
\small
$N_\sigma$ as a function of $\delta$ and $J_{CP}$ for the TBM and BM models 
in the standard ordering setup.  The dashed lines represent 
the results of the global fit reported in \cite{Fogli:2012ua} 
while the thick ones represent the results we obtain in our setup. 
Blue lines are for normal hierarchy while the red ones are 
for inverted hierarchy. These bounds are obtained minimizing 
the value of $N_\sigma$ in the parameter space for a 
fixed value of $\delta$ (left plots) or $J_{CP}$ (right plots).
}
\end{figure}
%

In Fig.~\ref{fig:results_delta} we show the contours of 
$N_\sigma = \sqrt{\chi^2}$ in the ($\sin^2 \theta_{23}, \delta$) 
plane, where the value of $\sin \theta_{13}$ has been marginalized. 
The blue dashed lines represent the contours of constant 
$J_{CP}$ (in units of $10^{-2}$).
In Figs.~\ref{fig:Nsigma_bounds_StOrd} and \ref{fig:bounds_Jcp}, 
starting from the same likelihood function, we show the 
bounds on the neutrino mixing parameters and $J_{CP}$ 
in each scheme, both for normal and inverted neutrino mass hierarchy.
These bounds are obtained minimizing the 
$\chi^2$ in the parameter 
space of the model, keeping as a constraint the 
value of the corresponding parameter.
To make a direct comparison of the bounds obtained 
in the scheme considered by us with the general 
bounds obtained in the global fit in \cite{Fogli:2012ua}, 
we show the results from \cite{Fogli:2012ua} with thin 
dashed lines.
Thus, the thin dashed lines in Fig.~\ref{fig:bounds_Jcp} are the 
bounds on $J_{CP}$ obtained using directly the 
results of the global fit \cite{Fogli:2012ua} 
and eq. \eqref{eq:Jcp_std}, and 
represent the present status of 
our knowledge on this observable assuming the 
standard $3$-neutrino mixing setup
\footnote{More refined bounds on $J_{CP}$ in the 
standard parametrisation of the PMNS 
matrix could be obtained by the authors of 
\cite{Fogli:2012ua}, using the full likelihood function.}.
The thick solid lines represent the results obtained 
in the scheme with standard ordering considered.
The blue and red color lines correspond respectively 
to  the cases of normal and inverted neutrino mass hierarchy; 
in the case when the two bounds are 
essentially identical we used purple color lines.

From Figs.~\ref{fig:results_delta} and \ref{fig:Nsigma_bounds_StOrd} 
we see that both the tribimaximal and bimaximal cases are well 
compatible with data. The $1\sigma$ difference between the minimum of 
$N_\sigma$ in the two cases is due to the fact that the 
bound on $\delta$ obtained in \cite{Fogli:2012ua} favours values of 
$\delta \sim \pi$ (see Table~\ref{tab:globalfit_PMNS_angles}), which is indeed the value needed in 
the bimaximal mixing (or LC) scheme to lower the value of 
$\theta_{12}$ from $\theta_{12}^\nu = \pi/4$, 
while the tri-bimaximal mixing scheme prefers 
$|\cos \delta| \ll 1$ (see Subsection~\ref{sec:StanOrd}). 

The results we obtain
for $\sin^2\theta_{12}$, $\sin^2\theta_{23}$ and $\sin^2\theta_{13}$ 
(i.e., the best fit values and the $3\sigma$ ranges)
in the case of tri-bimaximal mixing are similar to those given in 
\cite{Fogli:2012ua}. In contrast, our results for 
the Dirac phase $\delta$ and, correspondingly, for 
the $J_{CP}$ factor, are drastically different. 
For the best fit values and the $3\sigma$ allowed ranges
\footnote{These ranges are obtained imposing: 
$\sqrt{\Delta \chi^2} = \sqrt{N_\sigma^2 - (N_\sigma^{min})^2} \equiv 3$.} 
of $\delta$ and $J_{CP}$ we find
(see also Table~\ref{table:bfAND3sigma}):
%
%
\begin{table}[t]
\centering
\begin{tabular}{l l | c | c }
\toprule
		& 	& Best fit & 3$\sigma$ range  \\ \hline
		& $J_{CP}$ (NH)  & $-0.034$ & $-0.039 \div -0.028 \oplus 0.028 \div 0.039 $ \\
		& $J_{CP}$ (IH)  & $-0.034$ & $-0.039 \div -0.026 \oplus 0.027 \div 0.039 $ \\  
		& $\delta$ (NH)  & $4.64$ & $1.38 \div 1.97 \oplus 4.29 \div 4.91 $ \\  
	TBM	& $\delta$ (IH)  & $4.64$ & $1.39 \div 2.17 \oplus 4.04 \div 4.93 $ \\  
		& $\sin \theta_{13}$ & $0.16$ &  $0.13 \div 0.18$  \\ 
		& $\sin^2 \theta_{23}$ (NH) & $0.39$ &  $0.33 \div 0.64$  \\
		& $\sin^2 \theta_{23}$ (IH) & $0.39$ &  $0.34 \div 0.66$  \\ 
		& $\sin^2 \theta_{12}$ & $0.31$ &  $0.25 \div 0.36$  \\ \hline
		
		& $J_{CP}$ & $0.00$ & $-0.027 \div 0.026$ \\ 
		& $\delta$ (NH)  & $3.20$ & $2.35 \div 3.95 $ \\  
		& $\delta$ (IH)  & $3.27$ & $2.37 \div 3.94 $ \\  
	BM	& $\sin \theta_{13}$ & $0.16$ &  $0.13 \div 0.18$  \\ 
		& $\sin^2 \theta_{23}$ (NH) & $0.38$ &  $0.33 \div 0.47$  \\
		& $\sin^2 \theta_{23}$ (IH) & $0.39$ &  $0.34 \div 0.50$  \\ 
		& $\sin^2 \theta_{12}$ & $0.31$ &  $0.28 \div 0.36$  \\
\bottomrule
\end{tabular}
\caption{\label{table:bfAND3sigma} Best fit and 3$\sigma$ ranges (found fixing $\sqrt{\chi^2 - \chi^2_{min}} = 3$) in the standard ordering setup. When not explicitly indicated otherwise, the result applies both for normal hierarchy and inverted hierarchy of neutrino masses.}
\end{table}
%
\begin{eqnarray}
\label{eq:deltaSONHTBMpi2}
{\rm NH: } \quad \delta \cong 4.64\cong \frac{3\pi}{2}\,,& 1.38 \ltap \delta \ltap 1.97\,, &{\rm or}\\
\label{eq:deltaSONHTBM3pi2}
&  4.29 \ltap \delta \ltap 4.91\,,&\\
\label{eq:deltaSOIHTBMpi2}
{\rm IH: } \quad \delta \cong 4.64\cong \frac{3\pi}{2}\,,& 1.39 \ltap \delta \ltap 2.17\,, &{\rm or}\\
&  4.04 \ltap \delta \ltap 4.93\,,&
\label{eq:deltaSOIHTBM3pi2}
\end{eqnarray}
%
\begin{eqnarray}
\label{eq:JCPSONHTBMpi2}
{\rm NH: } \quad J_{CP} \cong -0.034\,,& 0.028 \ltap J_{CP} \ltap 0.039 \,,
&{\rm or}\\
\label{eq:JCPSONHTBM3pi2}
&  -0.039 \ltap J_{CP} \ltap -0.028\,,&\\
\label{eq:JCPSOIHTBMpi2}
{\rm IH: } \quad J_{CP} \cong -0.034\,,&  0.027 \ltap J_{CP} \ltap 0.039 \,,
&{\rm or}\\
&  -0.039 \ltap J_{CP} \ltap -0.026 \,.&
\label{eq:JCPSOIHTBM3pi2}
\end{eqnarray}
The $3\sigma$ intervals of allowed values of $\delta$ ($J_{CP}$) 
in eqs. (\ref{eq:deltaSONHTBMpi2}) and  (\ref{eq:deltaSOIHTBMpi2}) 
(eqs. (\ref{eq:JCPSONHTBMpi2}) and  (\ref{eq:JCPSOIHTBMpi2})) 
are associated with the local minimum at $\delta \cong \pi/2$ 
($J_{CP} \cong 0.034$) in Fig. \ref{fig:bounds_Jcp} 
upper left (right) panel, while those given in  
eqs. (\ref{eq:deltaSONHTBM3pi2}) and  
(\ref{eq:deltaSOIHTBM3pi2}) (eqs. (\ref{eq:JCPSONHTBM3pi2}) 
and  (\ref{eq:JCPSOIHTBM3pi2})) are related 
to the absolute minimum at 
$\delta \cong 3\pi/2$ ($J_{CP}\cong -0.034$). 

The results we have obtained, reported in
Figs.~\ref{fig:results_delta} and \ref{fig:bounds_Jcp}, 
and in eqs. (\ref{eq:deltaSONHTBMpi2}) - (\ref{eq:JCPSOIHTBM3pi2}), 
are quasi-degenerate with respect to 
$J_{CP} \rightarrow - J_{CP}$, 
or $\delta \rightarrow (2 \pi - \delta)$. 
This stems from the fact 
that the phase $\phi$ enters
into the expressions for the 
mixing angles only via its cosine, see eqs. 
(\ref{eq:s12sqNOBM1}) and (\ref{eq:s12sqNOTBM1}). 
This symmetry is slightly broken only by the explicit bound 
on $\delta$ given in \cite{Fogli:2012ua}, 
which is graphically represented in Fig.~\ref{fig:bounds_Jcp} 
by the asymmetry of the dashed lines 
showing that negative values of $J_{CP}$ are slightly favored.

As Figs.  \ref{fig:results_delta} and ~\ref{fig:bounds_Jcp} show,
in the case of tri-bimaximal mixing, 
the CP conserving values of $\delta =0;\pi;2\pi$ 
is excluded with respect to the best fit CP violating
values  $\delta \cong \pi/2;3\pi/2$ at more than $4\sigma$. 
Correspondingly,  $J_{CP} = 0$ is also excluded with respect 
to the best-fit values $J_{CP} \simeq (-0.034)$ and 
$J_{CP}\simeq 0.034$ at more than $4 \sigma$.
It follows from eqs. (\ref{eq:deltaSONHTBMpi2}) - (\ref{eq:JCPSOIHTBM3pi2}) 
(see also Table~\ref{table:bfAND3sigma})
that the $3\sigma$ allowed ranges of values 
of both $\delta$ and 
$J_{CP}$ form rather narrow intervals.
These are the most striking predictions of the 
scheme with standard ordering and tri-bimaximal mixing 
under investigation. 

We obtain different results assuming bimaximal mixing
in the neutrino sector. Although in this case 
the best fit values of  $\sin^2\theta_{12}$, $\sin^2\theta_{23}$, 
$\sin^2\theta_{13}$ and $\delta$ practically coincide with those 
found in \cite{Fogli:2012ua}, the $3\sigma$ allowed intervals 
of values of $\sin^2\theta_{12}$ and especially of 
$\sin^2\theta_{23}$ and $\delta$
differ significantly from 
those given in \cite{Fogli:2012ua}.

For the best fit values and the $3\sigma$ intervals of  
 $\sin^2\theta_{12}$ and $\sin^2\theta_{23}$ 
we get (see also  Table~\ref{table:bfAND3sigma}):

\begin{eqnarray}
\label{eq:s12sqNOBM3sig}
\sin^2\theta_{12} \cong 0.31\,,~~~0.28 \ltap \sin^2\theta_{12} \ltap 0.36\,; \\ [0.30cm]
\label{eq:s23sqNOBMNH3sig}
{\rm NH: } \quad \sin^2\theta_{23} \cong 0.38\,, ~~0.33\ltap \sin^2\theta_{23} \ltap 0.47\,; \\ [0.30cm]
{\rm IH: } \quad \sin^2\theta_{23} \cong 0.39\,, ~~0.34\ltap \sin^2\theta_{23} \ltap 0.50\,.
\label{eq:s23sqNOBMIH3sig}
\end{eqnarray}
%
As in \cite{Fogli:2012ua}, we find for the best fit value 
of $\delta$ and $J_{CP}$: $\delta \cong \pi$ and $J_{CP} \cong 0$.
However, the $3\sigma$ range of $\delta$ and, correspondingly, of 
$J_{CP}$, we obtain differ from those derived in  \cite{Fogli:2012ua}: 
\begin{eqnarray}
\label{eq:deltaJCPSONHBM3sigm}
	{\rm NH: } \quad 2.35 \ltap \delta \ltap 3.95 \,;~~~~ 
	-\,0.027 \ltap J_{CP} \ltap 0.026\,. \\
	{\rm IH: } \quad 2.37 \ltap \delta \ltap 3.94 \,;~~~~ 
	-\,0.027 \ltap J_{CP} \ltap 0.026\,. 
\label{eq:JCPSOIHBM3sigm}
\end{eqnarray}
%
  We see, in particular, that also in this case the 
Dirac CPV phase $\delta$ is constrained to 
lie in a narrow interval around the value $\delta \simeq \pi$. 
This and the constraint $\sin^2 \theta_{23} \lesssim 1/2$ 
are the most important predictions of the scheme 
with standard ordering and bimaximal neutrino mixing.

%
\section{Results with the Inverse Ordering}
\label{sec:inverse}
%

The case of inverse ordering is qualitatively and quantitatively 
different from the case of standard ordering. 
For given values of $\theta_{12}^\nu$, $\theta^\nu_{23}$, 
the number of parameters is the same as in the PMNS matrix. 
Still, not all values of $U$ can be obtained, as we shall see. 

The constraints on the reactor and atmospheric neutrino mixing 
angles are the same for bimaximal and tri-bimaximal 
mixing and can be derived directly from 
eq. (\ref{eq:chlep_corrections_inv_ord_BMandTBM}). 
For any given value of the phase $\psi$, 
any values of $\theta_{13}$ and $\theta_{23}$ in the ranges
\be 
\begin{split}
	\qquad 0 \leq \sin \theta_{13} \leq &\: \frac{1}{\sqrt{2}}, \\
	\qquad 0 \leq \sin^2 \theta_{23} \leq &\: \frac{\cos 2 \theta_{13}}{\cos^4 \theta_{13}} \simeq 1 + \mathcal{O}(\sin^2 \theta_{13}),
\end{split} 
\ee
%
can be obtained by an appropriate choice of 
$\omega^\prime$, $\theta^e_{12}$ and $\theta^e_{23}$. 
Clearly, the range of values allowed for $\theta_{13}$ and $\theta_{23}$ 
covers the full experimentally allowed range. 
The solar neutrino mixing angle can now be expressed in terms of 
$\theta_{13}$ and $\psi$ as in 
\eq{chlep_corrections_inv_ord_BMandTBM}. 
Any value of $\theta_{12}$ in the interval
\begin{align}
	\text{BM}_{IO}:& \nonumber \\
	\frac{1}{2} \frac{1 - 2 \sin \theta_{13} \sqrt{\cos 2 \theta_{13}} - \sin^2 \theta_{13}}{\cos^2 \theta_{13}} \leq  \sin^2 \theta_{12}& \leq  \frac{1}{2} \frac{1 + 2 \sin \theta_{13} \sqrt{\cos 2 \theta_{13}} - \sin^2 \theta_{13}}{\cos^2 \theta_{13}}, \\
	\text{TBM}_{IO}:& \nonumber \\
	\frac{1}{3} \frac{1 - 2 \sqrt{2} \sin \theta_{13} \sqrt{\cos 2 \theta_{13}} }{\cos^2 \theta_{13}} \leq  \sin^2 \theta_{12}& \leq  \frac{1}{3} \frac{1 + 2 \sqrt{2} \sin \theta_{13} \sqrt{\cos 2 \theta_{13}} }{\cos^2 \theta_{13}},
\end{align}
%
can then be obtained for an appropriate choice of $\psi$. 
At leading order in $\sin \theta_{13}$ these bounds become
\be 
\begin{split}
	\text{BM}_{IO}: & \quad \frac{1}{2} - \sin \theta_{13} \lesssim  \sin^2 \theta_{12} \lesssim  \frac{1}{2} + \sin \theta_{13}, \\
	\text{TBM}_{IO}: & \quad \frac{1}{3} - \frac{2 \sqrt{2}}{3} \sin \theta_{13} \lesssim \sin^2 \theta_{12} \lesssim  \frac{1}{3} + \frac{2 \sqrt{2}}{3} \sin \theta_{13}.
\label{eq:appr_bounds_solarangle_IO}
\end{split}
\ee
%
Given the experimental bounds on the PMNS 
angles found in the global fit \cite{Fogli:2012ua}, 
see Table~\ref{tab:globalfit_PMNS_angles}, 
one can immediately 
notice that while the tri-bimaximal case is perfectly 
compatible with the data, the bimaximal case has a 
$\sim 2 \sigma$ tension in the prediction of 
the solar neutrino mixing angle parameter $\sin^2\theta_{12}$.
\begin{figure}[p]
 \centering
 \vspace*{-1cm}
 \fbox{\footnotesize Inverse Ordering - Tribimaximal} \\[0.5cm]
 \vspace*{-0.2cm} 
 \includegraphics[height=6.5cm]{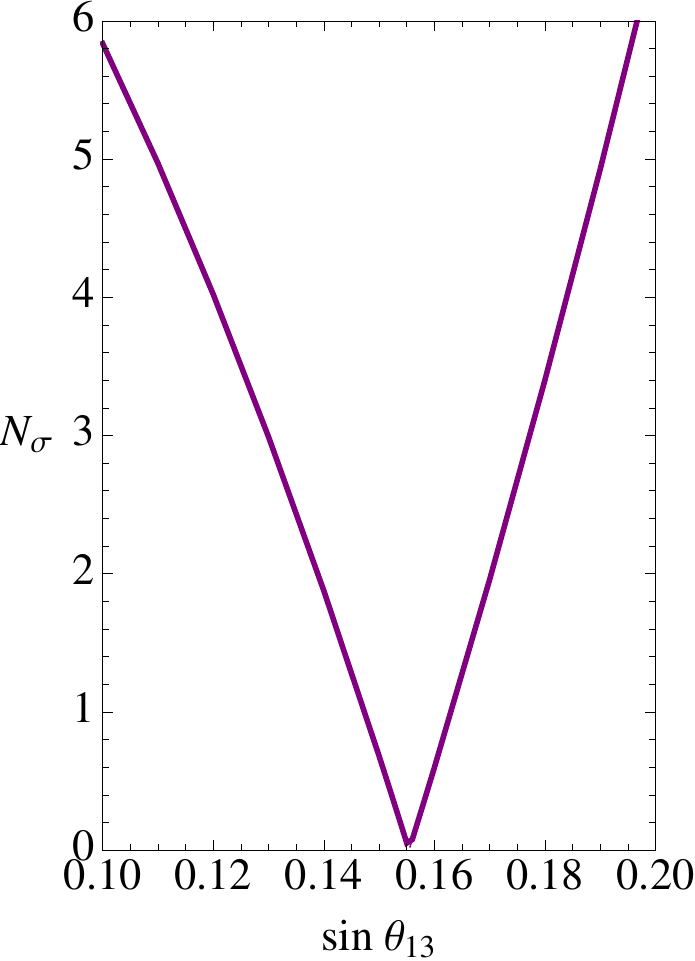} 
 \includegraphics[height=6.5cm]{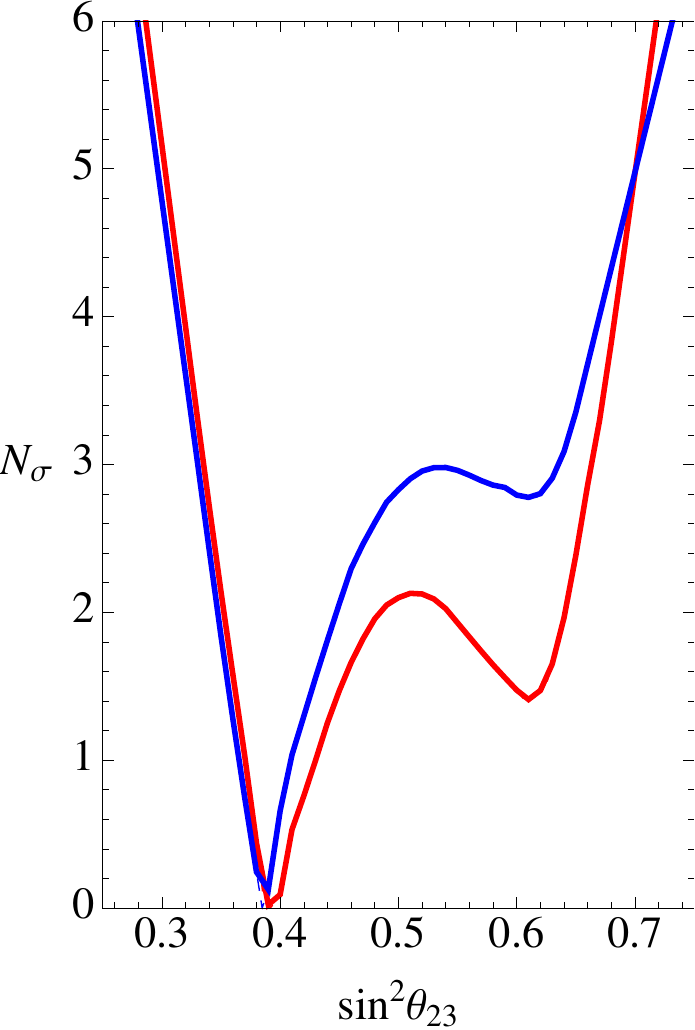}
 \includegraphics[height=6.5cm]{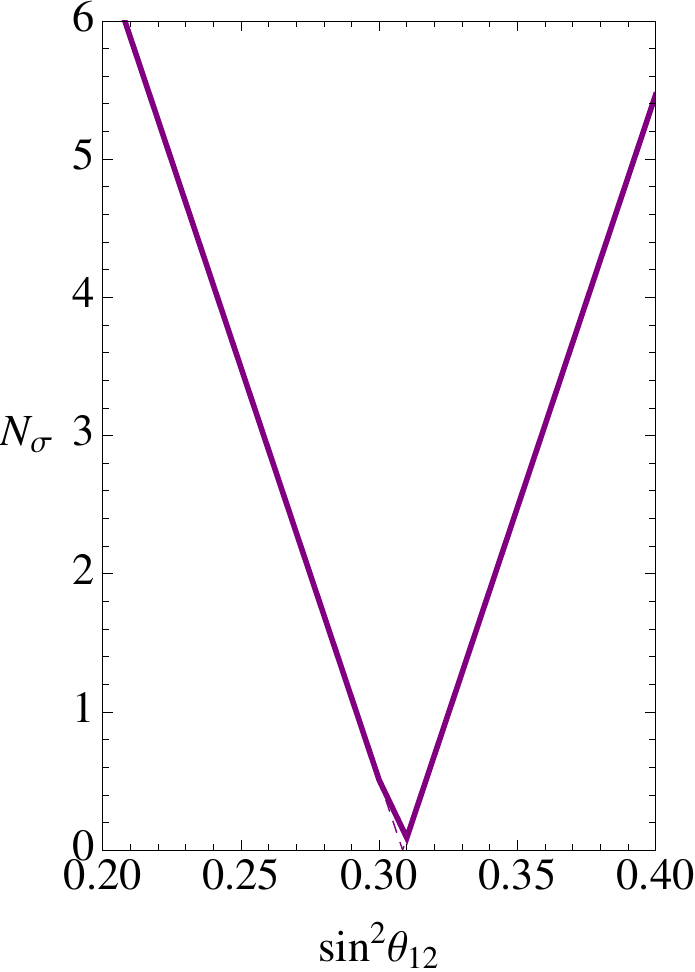}\\
 \vspace*{0.2cm}
 \fbox{\footnotesize Inverse Ordering - Bimaximal} \\[0.5cm]
 \vspace*{-0.2cm} 
 \includegraphics[height=6.5cm]{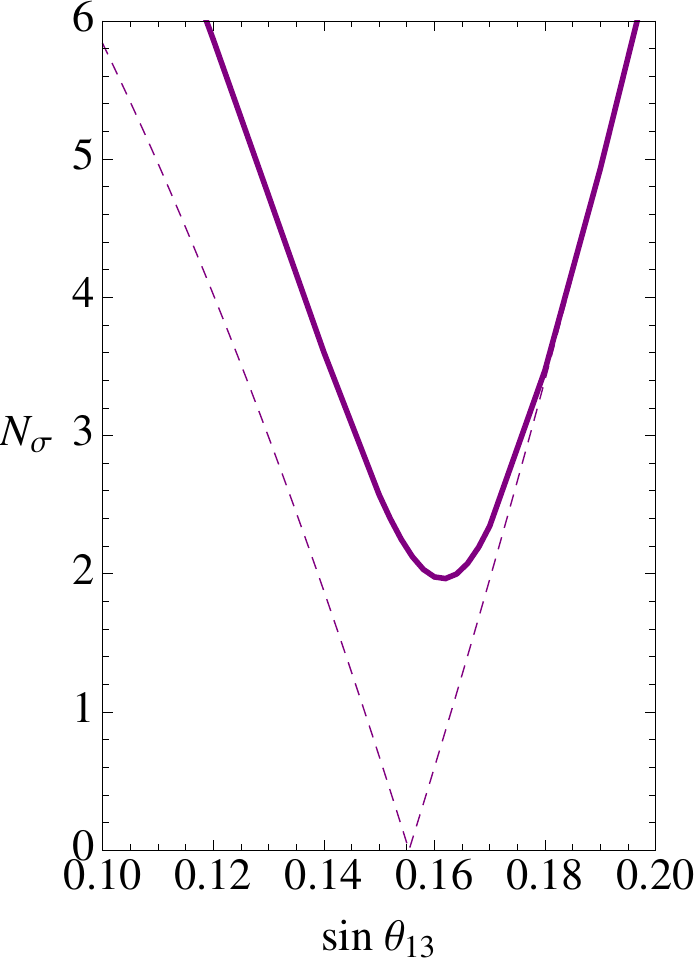} 
 \includegraphics[height=6.5cm]{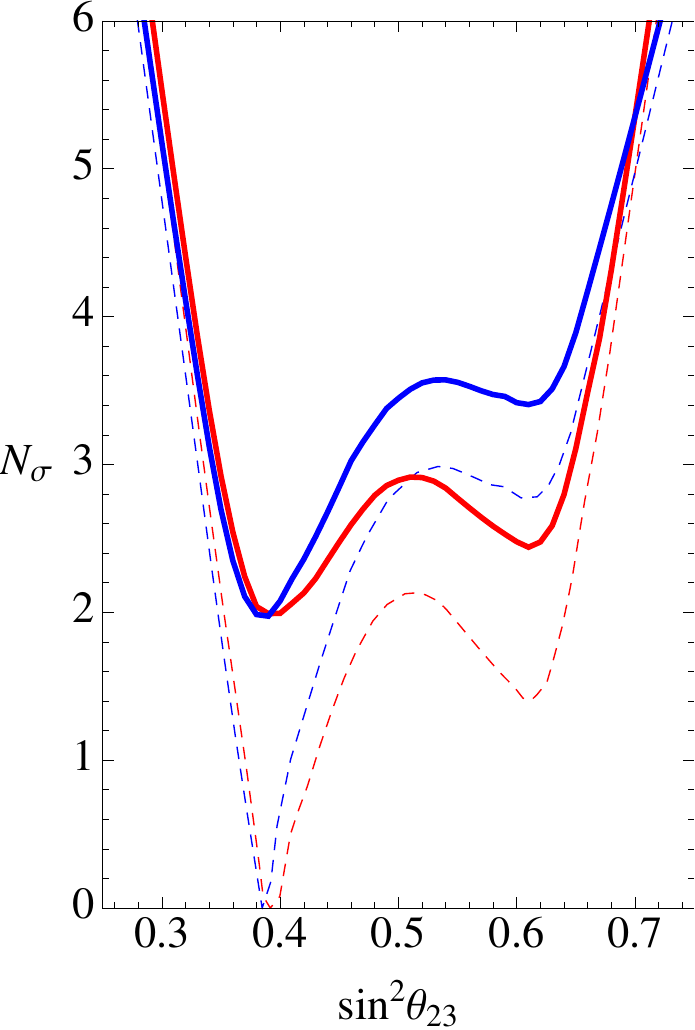}
 \includegraphics[height=6.5cm]{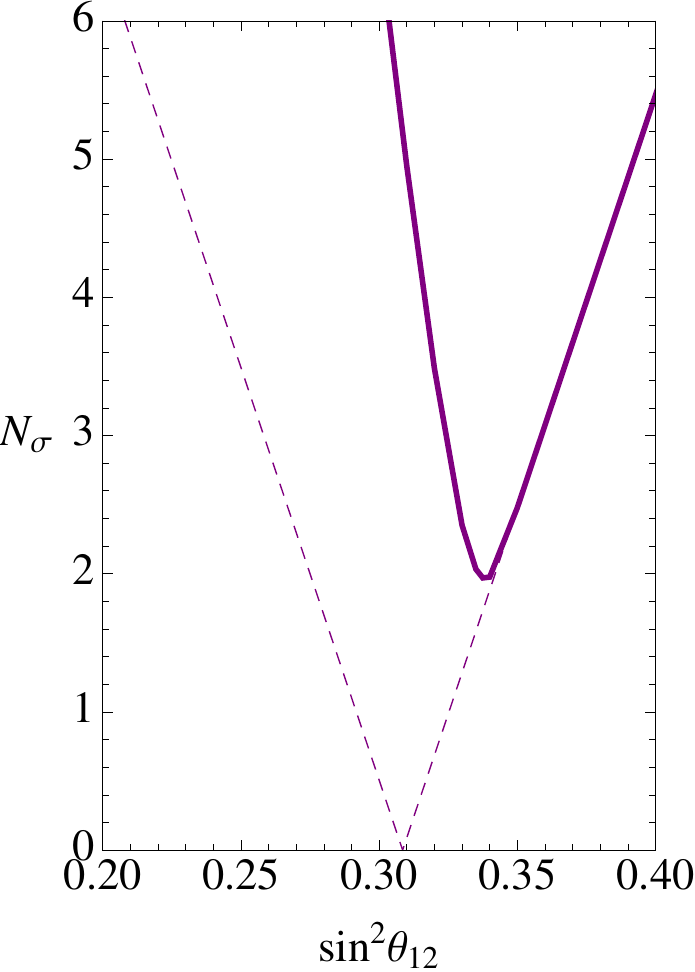}
 \caption{$N_\sigma$ as a function of each mixing angle for the TBM and 
BM models with the inverse ordering setup. The dashed lines represent the 
results of the global fit reported in \cite{Fogli:2012ua} while the thick ones 
represent the results we obtain in our setup. Blue lines 
are for normal hierarchy
while the red ones are for inverted hierarchy (we use purple when the two 
bounds are approximately identical). These bounds are obtained minimizing 
the value of $N_\sigma$ in the parameter space for fixed 
value of the showed mixing angle.}
\label{fig:Nsigma_bounds_IO}
\end{figure}

\begin{figure}[t]
\begin{center}
\fbox{\footnotesize Inverse Ordering} \\[0.5cm]
\hspace*{-0.65cm} 
\begin{minipage}{0.5\linewidth}
\begin{center}
	\hspace*{-0.8cm} 
	\fbox{\footnotesize Tribimaximal} \\[0.05cm]
	\includegraphics[height=65mm]{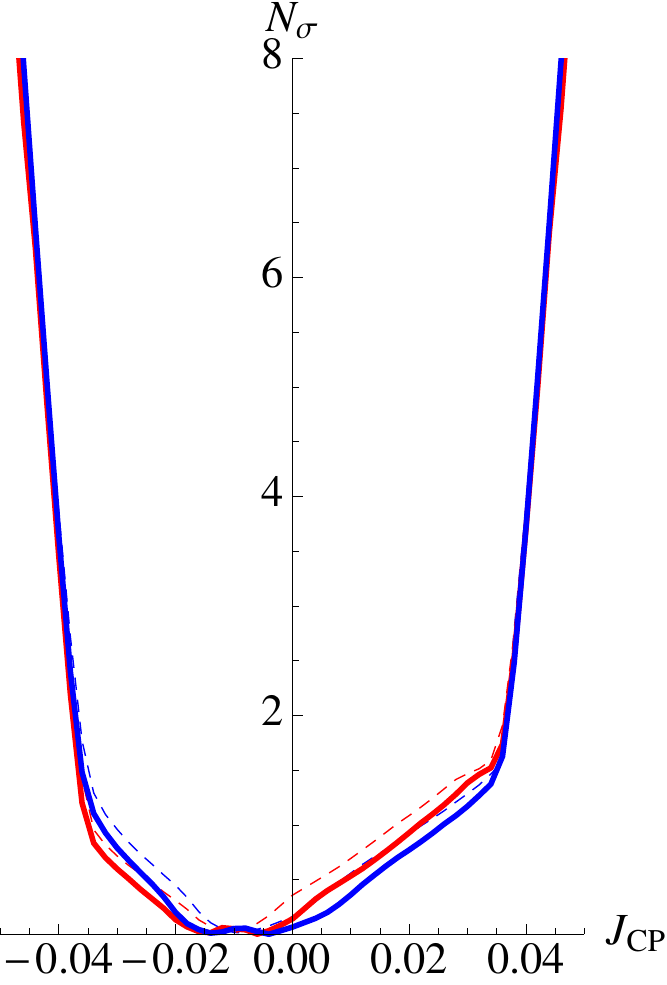}\\
	\mbox{\footnotesize (a)} \\
\end{center}
\end{minipage}
\hspace{0.25cm}
\begin{minipage}{0.5\linewidth}
\begin{center}
	\hspace*{-0.8cm}
	\fbox{\footnotesize Bimaximal} \\[0.05cm]
	\includegraphics[height=65mm]{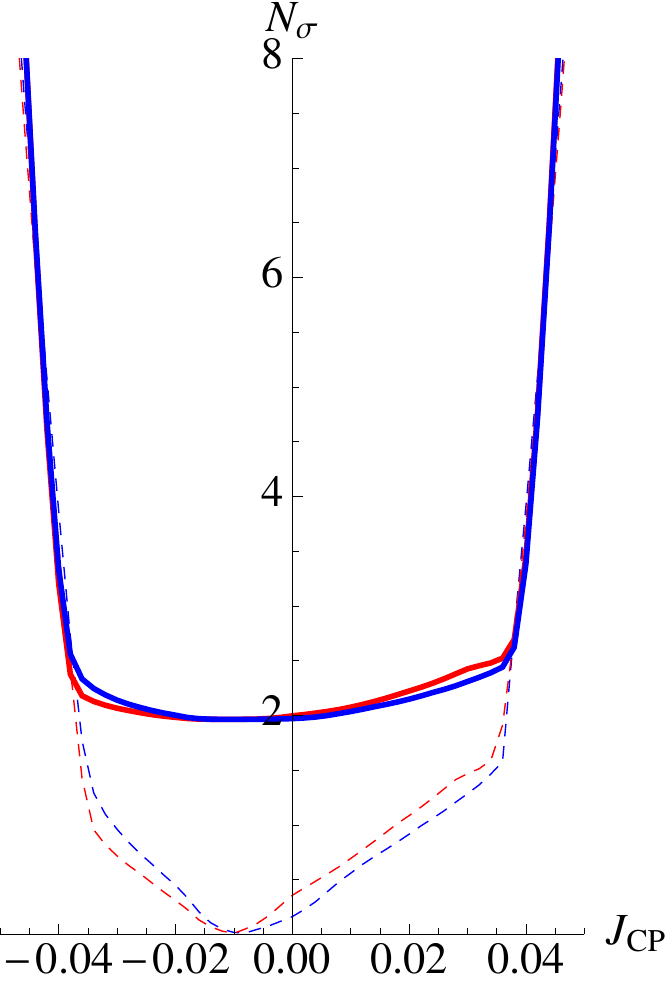}\\
	\mbox{\footnotesize (b)} \\
\end{center}
\end{minipage} 

\end{center}
\caption{\label{fig:bounds_Jcp_io} 
\small
$N_\sigma$ as a function of $J_{CP}$ for the TBM and BM models 
in the inverse ordering setup. The dashed lines represent 
the results of the global 
fit reported in \cite{Fogli:2012ua} while the thick ones represent the 
results we obtain in our setup. Blue lines are for normal 
neutrino mass hierarchy while 
the red ones are for inverted hierarchy.
These bounds are obtained minimizing the value 
of $N_\sigma$ in the parameter space for a fixed value of $J_{CP}$.
}
\end{figure}
%
As was done for the standard ordering case, 
we construct the likelihood function and the $\chi^2$ 
for both models as described in Appendix \ref{app:Stat}, 
exploiting the constraints on $\sin^2 \theta_{12}$, 
$\sin^2 \theta_{23}$, $\sin^2 \theta_{13}$ and on 
$\delta$ obtained in \cite{Fogli:2012ua}, and using in this case 
as parameters $\sin \theta_{13}$, $\sin \theta_{23}^e$ and the 
phases $\psi$ and $\omega$.
We show in Figs.~\ref{fig:Nsigma_bounds_IO} and 
\ref{fig:bounds_Jcp_io} the bounds on the neutrino mixing 
angles and the $J_{CP}$ factor both in the cases of 
bimaximal and tri-bimaximal mixing in the neutrino sector, 
and for normal and inverted neutrino mass hierarchy.

From Fig.~\ref{fig:Nsigma_bounds_IO}, we see that in the case 
of tribimaximal mixing (upper row), 
the intervals of allowed values of 
the PMNS mixing angles obtained in the model under discussion and 
in the global fit performed in \cite{Fogli:2012ua} coincide.
This is a consequence of the fact that the 
4D parameter space of the model considered completely overlaps with the 
experimentally allowed parameter space in the PMNS parametrisation 
and therefore it does not give any additional constraint. 
It is consistent with the analytic bounds reported above as well.

  In the case of bimaximal mixing instead 
(Fig.~\ref{fig:Nsigma_bounds_IO} lower row), 
only a portion of the relevant PMNS 
parameter space is reachable, a fact that 
is reflected in the bounds on $\sin^2\theta_{12}$
given in eq. \eqref{eq:appr_bounds_solarangle_IO}. Values of $\theta_{12}$ in the upper part of its present experimental range are favoured in this case.

In both cases of tri-bimaximal and bimaximal mixing 
from the neutrino sector, 
the bounds on $\sin^2\theta_{13}$ and  $\sin^2\theta_{12}$
corresponding to the normal and inverted neutrino mass hierarchy
are approximately identical, while they differ for the atmospheric 
neutrino mixing angle and for the $J_{CP}$ factor.

Considering the expressions for $J_{CP}$ in  
eqs.~(\ref{eq:chlep_corrections_inv_ord_BM}) and 
(\ref{eq:chlep_corrections_inv_ord_TBM}) and 
Fig.~\ref{fig:bounds_Jcp_io}, we see that within 
$\sim 1\sigma$ from the best-fit point, 
every value in the ranges
\be
\left| J_{CP}^{BM} \right| \lesssim \frac{\sin \theta_{13}^{+1\sigma}}
{4} \sim 0.04, \qquad
\left| J_{CP}^{TBM} \right| \lesssim \frac{\sin \theta_{13}^{+1\sigma}}
{3 \sqrt{2}} \sim 0.038,
\ee
%
is allowed, where we have used the $1 \sigma$ upper bound on 
$\sin \theta_{13}$ from  Table~\ref{tab:globalfit_PMNS_angles}. 
As a consequence, we cannot make more specific 
predictions about the CP violation due to the 
Dirac phases $\delta$ in this case.
This is an important difference with respect to 
the standard ordering scheme
where, in the tri-bimaximal mixing case, 
relatively large values of the $|J_{CP}|$ factor 
lying in a narrow interval are predicted at 
$3\sigma$ and, in the bimaximal mixing case, 
$\delta$ is predicted to lie at 
$3 \sigma$ in a narrow interval around the value 
of $\delta \sim \pi$.

\newpage
%
\section{Summary and Conclusions}
\label{sec:Conclusions}
%
%
In this paper we considered the possibility that the 
neutrino mixing angle $\theta_{13}$ arises from the interplay 
of  12 and 23 rotations in the neutrino ($U_\nu$) and 
charged lepton ($U_e$) contributions to the PMNS neutrino mixing matrix 
($U = U^\dagger_e U_{\nu}$). We generalized previous 
work~\cite{Marzocca:2011} in two directions. 
First, we considered two possible orderings of 12 and 23 rotations 
in $U_e$, the ``standard'', $U_e\sim R^e_{23} R^e_{12}$, and 
the  ``inverse'', $U_e\sim R^e_{12} R^e_{23}$, while keeping 
the standard ordering in the neutrino sector, 
$U_\nu \sim R^\nu_{23} R^\nu_{12}$. Second,
in order to be able to accommodate 
a possible deviation of the atmospheric neutrino mixing angle
$\theta_{23}$ from $\pi/4$, 
we allowed the charged lepton 23 rotation angle 
(and possibly the neutrino one, in the standard case) 
to assume arbitrary values.
We considered the cases in which $U_\nu$ is in the 
bimaximal or tri-bimaximal form, or in the form resulting 
from the conservation of the lepton charge $L_e - L_\mu - L_{\tau}$ (LC). 
We took, of course, all relevant physical CP violation (CPV) phases
into account.

The case of normal ordering turns out to be particularly interesting. 
The  PMNS matrix can be parameterized in terms of the charged lepton 
and neutrino 12 rotation angles, $\theta^e_{12}$ and $\theta^\nu_{12}$, 
an effective 23 rotation angle, $\hat\theta_{23}\approx \theta_{23}$, 
and a CPV phase $\phi$. 
Once $\theta^\nu_{12}$ is fixed to the bimaximal 
(LC) or tri-bimaximal value, 
the number of parameters reduces to three, and 
the Dirac phase $\delta$ in the PMNS matrix
can be predicted in terms of the PMNS 
solar, atmospheric and reactor neutrino mixing
angles $\theta_{12}$, $\theta_{23}$ and $\theta_{13}$. 
Moreover, the range of possible values of the PMNS angles 
turns out to be constrained. 

  In the tri-bimaximal case, the  Dirac CPV phase $\delta$ 
is predicted to 
have a value $\delta\approx \pi/2$ or $\delta \approx 3\pi/2$, 
implying nearly maximal CP violation in neutrino oscillations,
while in the bimaximal (and LC) case 
we find  $\delta \approx \pi$ and, consequently, 
the CP violation effects in neutrino oscillations are 
predicted to be small.
The present data have a mild preference for the latter option
(see Table \ref{tab:globalfit_PMNS_angles} and, 
e.g.,  Fig. \ref{fig:bounds_Jcp}).
Moreover, $\theta_{23}$ 
is predicted to be below $\pi/4$ in 
the bimaximal case, which is also in agreement with 
the indications from the current global neutrino 
oscillations data. In the set-up considered by us,
the $\theta_{23} > \pi/4$ solution 
of the global fit analysis in~\cite{Fogli:2012ua} 
is disfavored. 

   The case of inverse ordering is qualitatively 
and quantitatively very  different. 
Fixing $U_\nu$ to the bimaximal or tri-bimaximal form is not 
sufficient to obtain a prediction: the number of free physical 
parameters in this case is four -- 
two angles and two CPV phases. Still, not all 
values of the four 
physical parameters in the PMNS matrix, 
$\theta_{12}$, $\theta_{23}$, $\theta_{13}$ and $\delta$,
can be reached in this parameterization. 
In the tri-bimaximal case, the ranges of parameters 
that can be reached overlaps with the experimental ranges, 
so that no predictions can be made. In the bimaximal case, however, 
this is not the case. One obtains, in fact, the approximate relation 
$\sin^2\theta_{12}\gtrsim 1/2-\sin\theta_{13}$, which is barely compatible 
with the data. As a consequence, i) there is a tension in the above 
relation that worsen the quality of the fit, and 
ii) values of $\theta_{12}$ in the upper part of its present 
experimental range are preferred. 
In both cases, no predictions for the Dirac CPV phase $\delta$ 
can be made. 
We did not consider here the LC case as 
it involves, in general, five parameters, 
while its ``minimal'' version, corresponding to 
setting $\theta^e_{23} = 0$, 
is equivalent to the standard ordering case with BM mixing 
(i.e., with $\theta^\nu_{12} = \pi/4$).

The fact that the value of the Dirac CPV phase 
$\delta$ is determined (up to an ambiguity of 
the sign of $\sin\delta$) by the values of the
three PMNS mixing angles,  $\theta_{12}$, $\theta_{23}$ 
and $\theta_{13}$, eqs. (\ref{2cosdNOpi4}) and  (\ref{2cosdNOsqrt3}),
are the most striking predictions of the 
scheme considered with standard ordering 
and bimaximal (LC) and tri-bimaximal mixing 
in the neutrino sector. 
As we have already indicated,
for the best fit values of $\theta_{12}$, 
$\theta_{23}$ and $\theta_{13}$ we get 
$\delta \cong \pi$ and $\delta \cong \pi/2$ or $3\pi/2$
in the cases of bimaximal and tri-bimaximal mixing, 
respectively. These results imply also that 
in the scheme with standard ordering we have discussed,  
the $J_{CP}$ factor which determines the magnitude 
of CP violation in neutrino oscillations, 
is also a function of the three mixing angles:
$J_{CP} = J_{CP}(\theta_{12},\theta_{23},\theta_{13}, 
\delta(\theta_{12},\theta_{23},\theta_{13})) = 
J_{CP}(\theta_{12},\theta_{23},\theta_{13})$.
This allowed us to obtain predictions for the range of 
possible values of $J_{CP}$ using the current data on 
$\sin^2\theta_{12}$, $\sin^2\theta_{23}$ 
and $\sin\theta_{13}$, which are given in 
eqs.   (\ref{eq:deltaSONHTBMpi2}) - (\ref{eq:JCPSOIHTBMpi2}) 
and eqs. (\ref{eq:deltaJCPSONHBM3sigm}) - (\ref{eq:JCPSOIHBM3sigm}).

 The predictions for $\sin^2\theta_{23}$, 
and for $\delta$ and $J_{CP}$ we have obtained 
in the scheme with standard ordering 
and bimaximal (or LC) or tri-bimaximal form of 
$U_{\nu}$ will be tested by the neutrino 
oscillation experiments able to determine whether 
$\sin^2\theta_{23} \lesssim 0.5$ or 
$\sin^2\theta_{23} > 0.5$,
and in the experiments
searching for CP violation 
in neutrino oscillations.

%
\section*{Aknowledgements}
This work was supported in part by the INFN program on
``Astroparticle Physics'', the Italian MIUR program on
``Neutrinos, Dark Matter and  Dark Energy in the Era of LHC''
the World Premier International Research Center 
Initiative (WPI Initiative), MEXT, Japan  (S.T.P.), 
by the European Union FP7-ITN INVISIBLES and UNILHC
(Marie Curie Action, PITAN-GA-2011-289442 and PITN-GA-2009-23792), 
and by the ERC Advanced Grant no. 267985 ``DaMESyFla''. 
Part of the work was done at the Galileo 
Galilei Institute for Theoretical Physics, 
Florence, Italy, which we thank for the kind 
hospitality and support.
%

%
\appendix
\section*{Appendix}
%

%
\section{Parametrisation of the PMNS matrix}
\label{app:Upmns_param}
%

 In the present  Appendix we show how the 
parametrisation of \eq{Upmns_param} follows from 
the ones in \eqs{Unu} and~(\ref{eq:Ue}). 
We start by writing explicitly the PMNS matrix  as
\be
	U = \Phi^*_e R_{12}(\theta^e_{12}) R_{23}(\theta_{23}^e) \Psi R_{23}(\theta_{23}^\nu) R_{12}(\theta_{12}^\nu) \Phi_\nu, 
\ee
%
where $\Psi = \diag(1, e^{i \psi}, e^{i \omega})$, 
without loss of generality. 
Any $2\times 2$ unitary matrix $V$ can be recast in 
the form $V = P R(\theta) Q$, 
where $P = \diag(e^{i\phi_1}, e^{i\phi_2})$, 
$Q = \diag(1, e^{i\omega_2})$ and 
$R(\theta)$ is a $2\times 2$ rotation. 
We use this to write
\be
	R_{23}(\theta_{23}^e) \Psi R_{23}(\theta_{23}^\nu) = \Phi^\prime R_{23}(\hat{\theta}_{23}) \Omega,
\ee
%
where $R_{23}(\hat{\theta}_{23})$ is an orthogonal rotation in the 23 block with
\be
	\sin \hat{\theta}_{23} = \left| \cos \theta_{23}^e \sin \theta_{23}^\nu + e^{i(\omega - \psi)} \sin \theta_{23}^e \cos \theta_{23}^\nu  \right|,
	\label{eq:theta23hat_relation}
\ee
%
$\Phi^\prime = \diag(e^{i\phi_1}, e^{i\phi_2}, e^{i\phi_3})$, 
and $\Omega = \diag(1, e^{i\omega_2}, e^{i\omega_3})$. 
An explicit solution for the angles in terms 
of the original parameters is
\be 
\begin{split}
	&\phi_1 = 0, \qquad \phi_2 = \delta_c + \delta_s + \psi - \omega, \qquad \phi_3 = 0, \\
	& \omega_2 = -  \delta_s + \omega, \qquad \omega_3 = -  \delta_c + \omega,
\end{split} 
\ee
%
where
\be \begin{split}
	\delta_s &= \text{Arg}\left( \cos \theta_{23}^e \sin \theta_{23}^\nu + e^{i(\omega - \psi)} \sin \theta_{23}^e \cos \theta_{23}^\nu \right), \\
	\delta_c &= \text{Arg}\left( \cos \theta_{23}^e \cos \theta_{23}^\nu - e^{i(\omega - \psi)} \sin \theta_{23}^e \sin \theta_{23}^\nu \right).
\end{split} 
\ee
%
Considering now also the $R_{12}(\theta_{12}^\nu)$ rotation, we obtain
\be
	R_{23}(\hat{\theta}_{23}) \Omega R_{12}(\theta_{12}^\nu) = \Phi^{\prime \prime} R_{23}(\hat{\theta}_{23}) R_{12}(\theta_{12}^\nu) \: Q^{\prime \prime},
\ee
%
with $\Phi^{\prime \prime} = \diag(1, e^{i\omega_2}, e^{i \omega_2})$ 
and $Q^{\prime \prime} = \diag(1, 1, e^{i(\omega_3 - \omega_2)})$.
The phases in $Q^{\prime \prime}$ add to the ones 
in $Q^\prime$ and are Majorana phases. 
The ones in $\Phi^{\prime \prime}$, instead, 
add to the ones in $\Phi^\prime$:
\be
	\Phi^{\prime} \Phi^{\prime \prime} = e^{i\phi_1} \diag(1, e^{i(\phi_2 - \phi_1 + \omega_2)}, e^{i(\phi_3 - \phi_1 + \omega_2)}).
\ee
%
The phase in the 33 position commutes with 
$R_{12}(\theta_{12}^e)$. Together with the overall 
phase $\phi_1$, it will describe the unphysical 
phase matrix $P$ in \eq{Upmns_param}:
\be
	P = e^{i\phi_1} \diag(1, 1, e^{i(\phi_3 - \phi_1 + \omega_2 )}).
\ee
%
We see that the only physical Dirac CP violating phase 
in this parametrisation is contained in the matrix 
$\Phi = \diag(1, e^{i\phi}, 1)$, with
\be
\phi = \phi_2 - \phi_1 + \omega_2 = \psi + \delta_c.
\ee

%
\section{Statistical analysis}
\label{app:Stat}
%

\begin{figure}
 \centering
 \includegraphics[height=5cm]{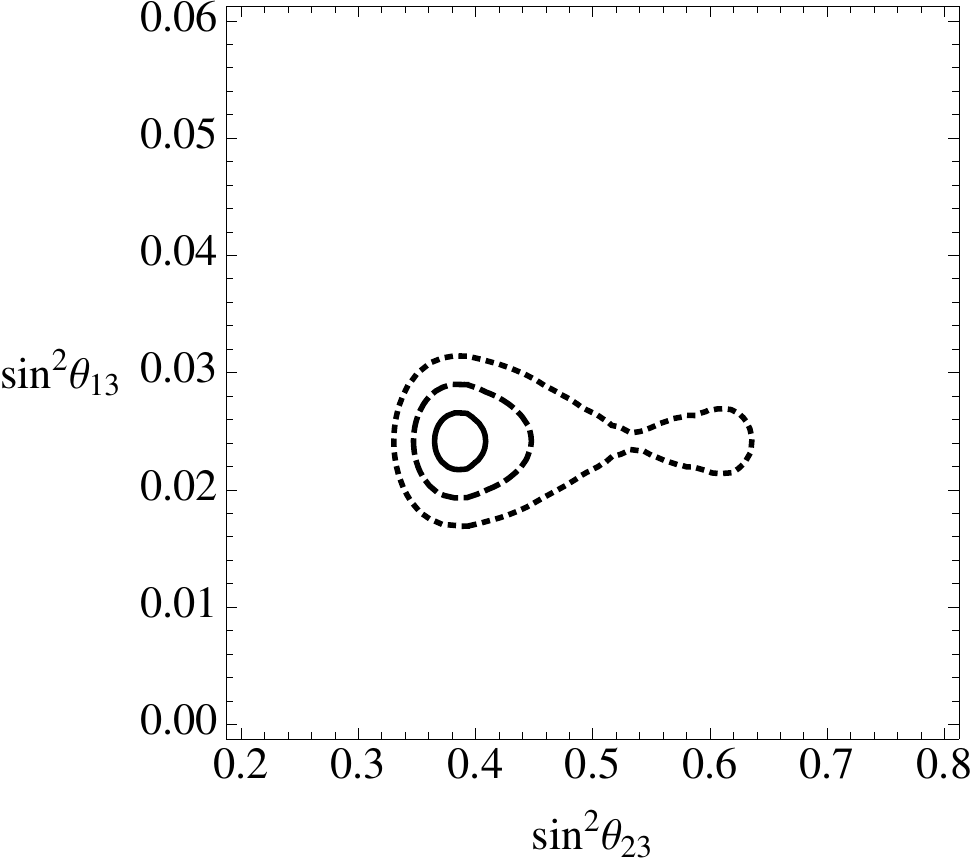} \hspace{1cm}
 \includegraphics[height=5cm]{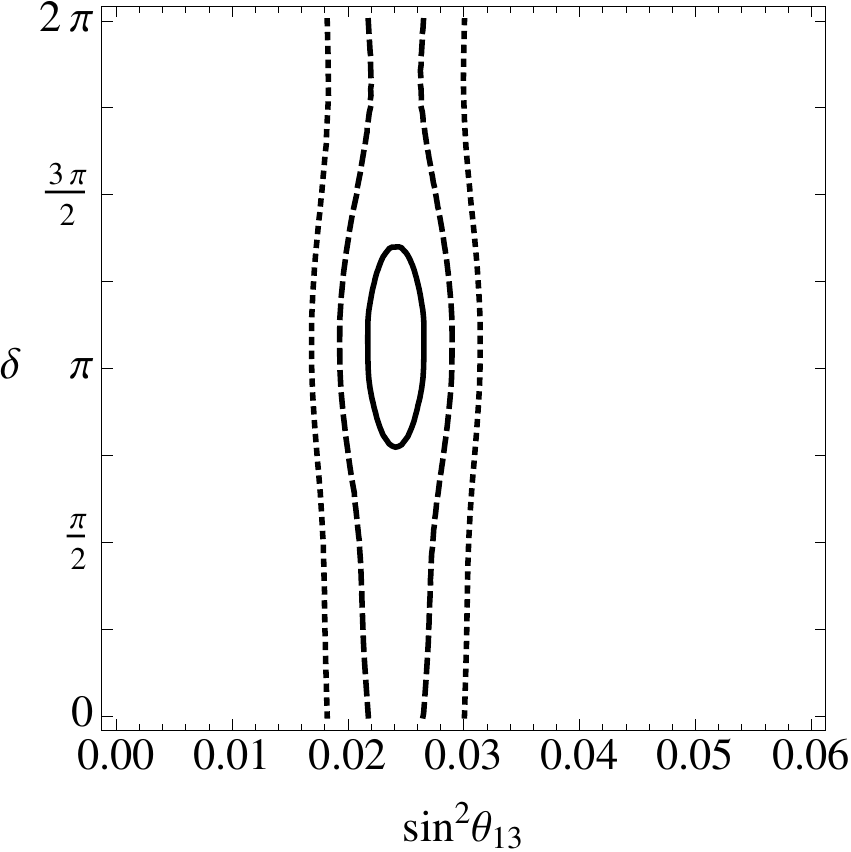}
 \caption{$1\sigma, 2\sigma, 3\sigma$ contours (respectively solid, dashed and dotted lines) of our global likelihood function in the $(\sin^2 \theta_{23}, \sin^2 \theta_{13})$ plane (left) and $(\sin^2 \theta_{13}, \delta)$ plane (right), using the data for NH. These plots can be compared with Fig.1 and Fig.2 of \cite{Fogli:2012ua} for NH. Undisplayed variables have been marginalized.}
\label{fig:correlations}
\end{figure}

In this appendix we describe the simplified statistical 
analysis performed to obtain the results. Our aim is to 
use the results of the global fit performed 
in \cite{Fogli:2012ua} to assess how well each of 
the models introduced in the previous section can 
fit the data. In particular, we use the constraints 
on the PMNS angles $\theta_{13}, \theta_{12}, \theta_{23}$ 
and on the phase $\delta$ for the normal hierarchy (NH) and inverted hierarchy (IH) cases, 
as derived in \cite{Fogli:2012ua}. There, the results 
are reported by plotting the value of 
$N_\sigma \equiv \sqrt{\Delta \chi^2}$ 
(with $\Delta \chi^2 = \chi^2 - \chi^2_{min}$) 
as a function of each observable, with the 
remaining ones marginalized away. We construct an approximate
global likelihood from these functions as
\be
	L_j(\alpha_j) = \exp\left(- \frac{\Delta \chi_{j}^2(\alpha_j)}{2} \right), \qquad L(\vec \alpha) = \prod_j^n L_j(\alpha_j),
\ee
%
where $\vec \alpha = \{ \sin^2 \theta_{13}, 
\sin^2 \theta_{23}, \sin^2 \theta_{12}, \delta \}$ 
are the observables relevant 
for our analysis, and we define
\be
	\chi^2(\vec \alpha) \equiv - 2 \log L(\vec \alpha)
\ee
%
and $N_\sigma(\vec \alpha) = \sqrt{\chi^2(\vec \alpha) } $.
In using this procedure 
we loose any information about possible correlations 
between different observables. The effect of this 
loss of information is however negligible, as one 
can check comparing our 1$\sigma$, 2$\sigma$ and 
3$\sigma$ contours in the $(\sin^2 \theta_{23}, \sin^2 \theta_{13})$ 
and $(\sin^2 \theta_{13}, \delta)$ planes shown 
in Figure \ref{fig:correlations} with the ones 
in Fig.~1 and Fig.~2 of \cite{Fogli:2012ua}.
%

Each model introduced in the previous section 
(which we dub with an index $m$) 
depends on a set of parameters 
${\bf x}^m = \{ x^m_i \}$, which are related to 
the observables via expressions 
$\alpha_j = \alpha_j^m({\bf x}^m)$, 
obtained from \eqs{chlep_corrections}, \eqref{eq:chlep_corrections_inv_ord}.
We then construct the likelihood function in the space 
of the parameters ${\bf x}^m$ as
\be
	L^m({\bf x}^m) = L(\vec \alpha^m ({\bf x}^m)).
\ee
%
We define $\chi^2 ({\bf x}^m) = - 2 \log L^m ({\bf x}^m) $ 
and $N_\sigma({\bf x}^m) = \sqrt{\chi^2({\bf x}^m)}$.
The last one is the function we use to produce the plots shown in Figures 2-6.
Finally, to obtain the best-fit point we use the maximum 
likelihood method.



\newpage
\section*{Addendum: Analysis with the 2013 Data}\label{Addendum}

We update our analysis 
using the results of the global fit performed in
\cite{Capozzi:2013csa}, in which the latest 2013 data 
from the Daya Bay, RENO, T2K and MINOS experiments are included.
A global fit including the indicated 2013 data 
was performed also in \cite{CGGMSchw12update} in which
the authors report similar results.
We restrict this update to the standard ordering case since 
the main conclusions for the inverse ordering one remain unchanged.
In this new global fit the authors find, for the Dirac CP violating phase, 
the best fit value of $\delta \cong 3\pi/2$. 
The CP conserving values $0$ and $\pi$
are disfavored, respectively, at $1.7\sigma$ and $1.5\sigma$ ($2\sigma$ and $1\sigma$)
in the case of NH (IH) neutrino mass spectrum, see Fig.~3 of \cite{Capozzi:2013csa}.
Another relevant difference of the results of this fit with respect 
to the previously obtained one in
\cite{Fogli:2012ua} is that the hint for atmospheric angle in the first 
octant is now  somewhat weaker than before: the best-fit value, located  
in the first octant, moved from $\sin^2 \theta_{23} \simeq 0.39$ to 
$\sin^2 \theta_{23} \simeq 0.43$, and the $\sqrt{\Delta \chi^2}$ of the 
relative minimum in the second octant dicreased
from $\sim 3\sigma$ ($\sim 1.5\sigma$) 
to $\sim 2.2\sigma$ ($\sim 0.5\sigma$) for NH (IH) spectrum.

\begin{figure}[p]
\begin{center}
\vspace*{-1.5cm}
\fbox{\footnotesize Standard Ordering - Normal Hierarchy} \\[0.5cm]
\hspace*{-0.65cm} 
\begin{minipage}{0.5\linewidth}
\begin{center}
	\hspace*{1cm} 
	\fbox{\footnotesize Tribimaximal} \\[0.05cm]
	\includegraphics[width=75mm]{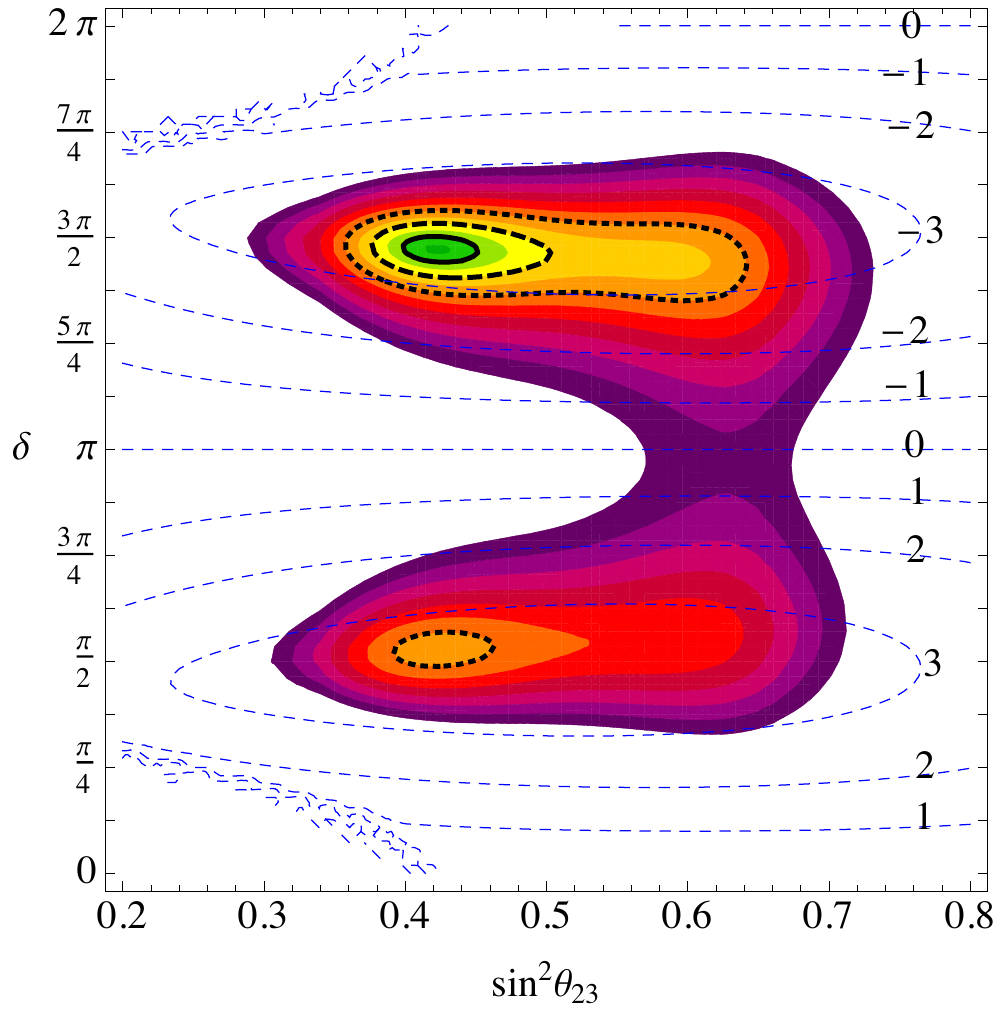}\\
	\vspace*{-0.2cm}
	\mbox{\footnotesize (a)} \\
\end{center}
\end{minipage}
\hspace{0.25cm}
\begin{minipage}{0.5\linewidth}
\begin{center}
	\hspace*{1cm}
	\fbox{\footnotesize Bimaximal} \\[0.05cm]
	\includegraphics[width=75mm]{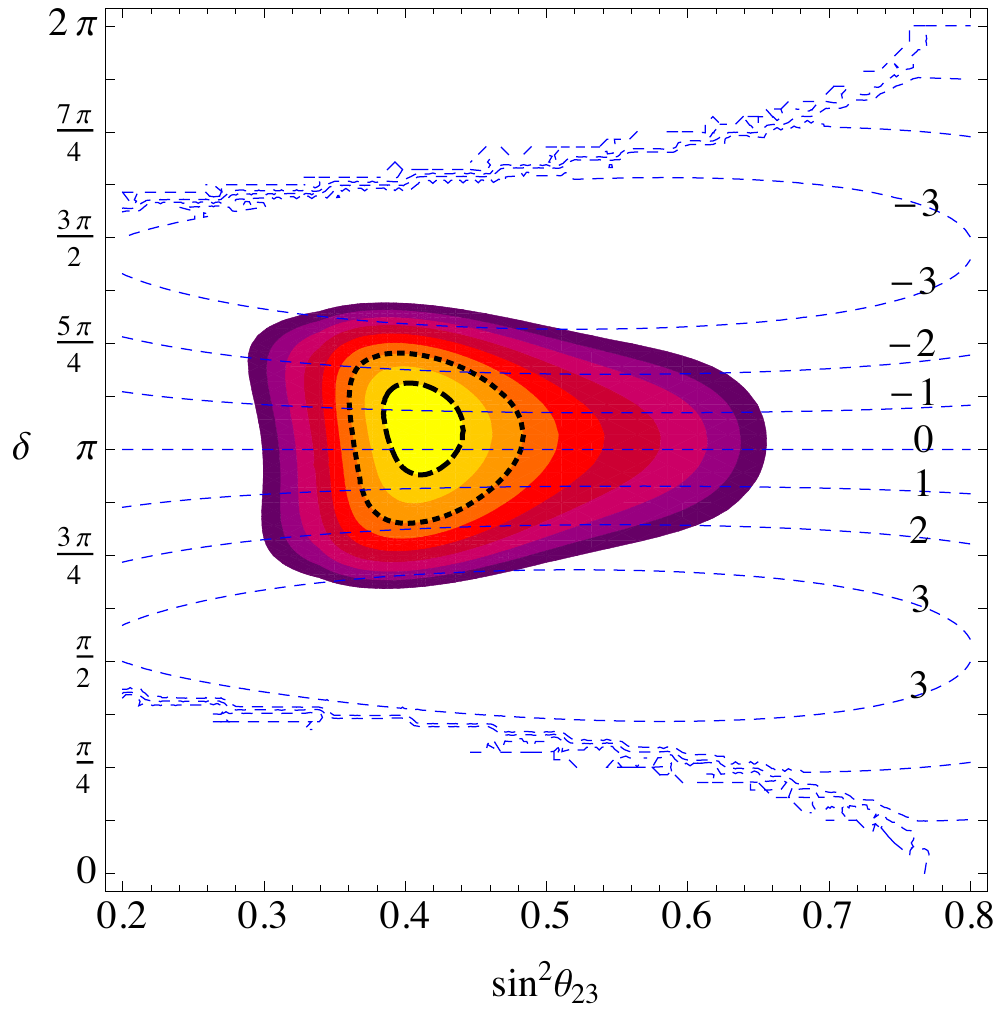}\\
	\vspace*{-0.2cm}
	\mbox{\footnotesize (b)} \\
\end{center}
\end{minipage} 
\\
\begin{minipage}{0.5\linewidth}
\begin{center}
	\includegraphics[width=75mm]{plots/legend_h_GYR}\\
\end{center}
\end{minipage}
\end{center}
\vspace*{-0.5cm}
\caption{\label{fig:results_delta_2013} 
\small
Contour plots for $N_\sigma = \sqrt{\chi^2}$
in the standard ordering setup and normal hierarchy of neutrino masses.
The value of the reactor angle $\theta_{13}$ has been marginalized. 
The solid, dashed and dotted thick lines represent respectively the 
$1\sigma, 2\sigma$ and $3\sigma$ contours. The dashed blue 
lines are contours of constant $\left| J_{CP} \right|$ 
in units of $10^{-2}$.}
\end{figure}

\begin{figure}[p]
 \centering
 \vspace*{-1cm}
 \fbox{\footnotesize Standard Ordering - Tribimaximal} \\[0.5cm]
 \vspace*{-0.2cm} 
 \includegraphics[height=6.5cm]{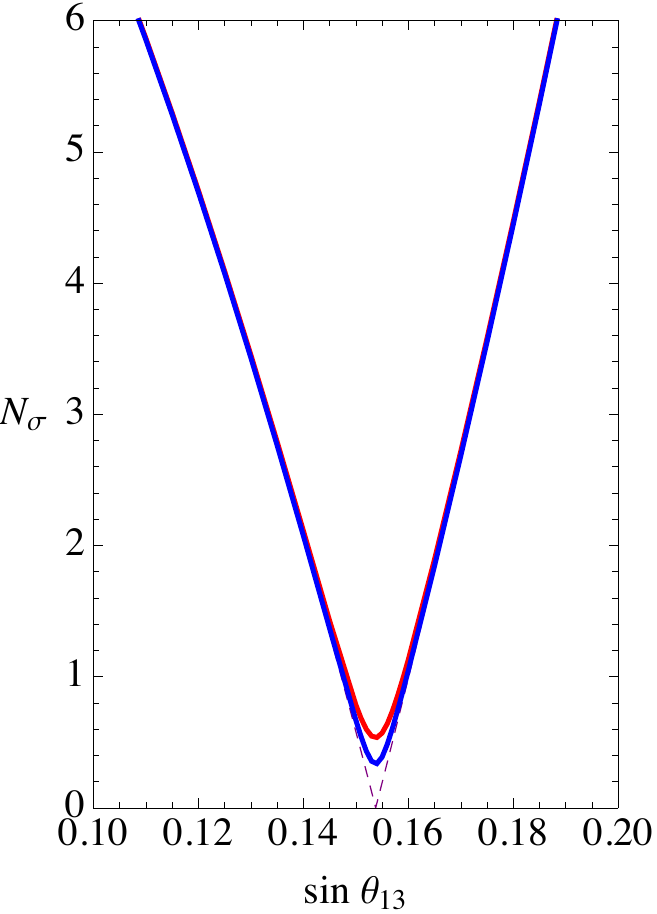} 
 \includegraphics[height=6.5cm]{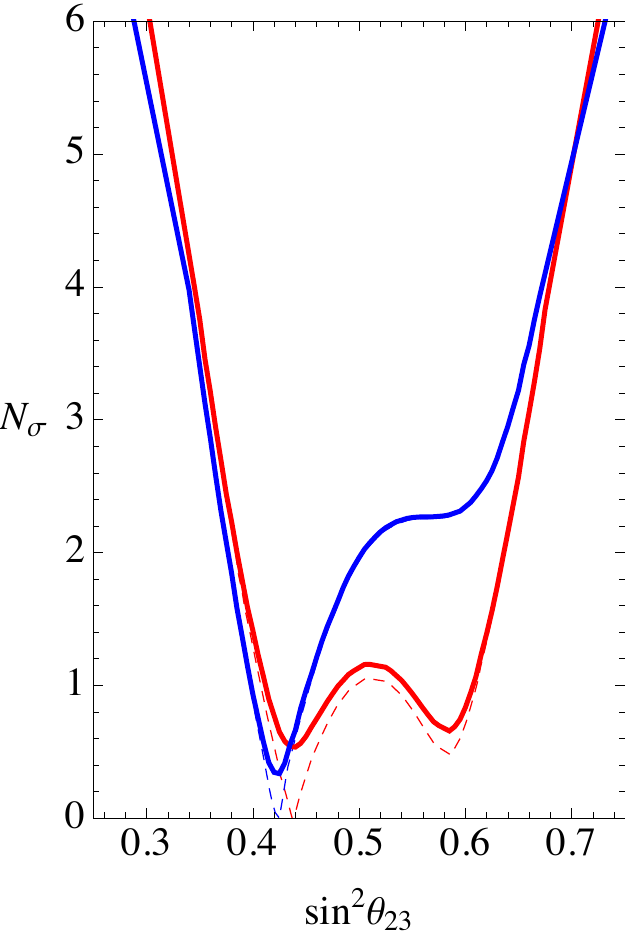}
 \includegraphics[height=6.5cm]{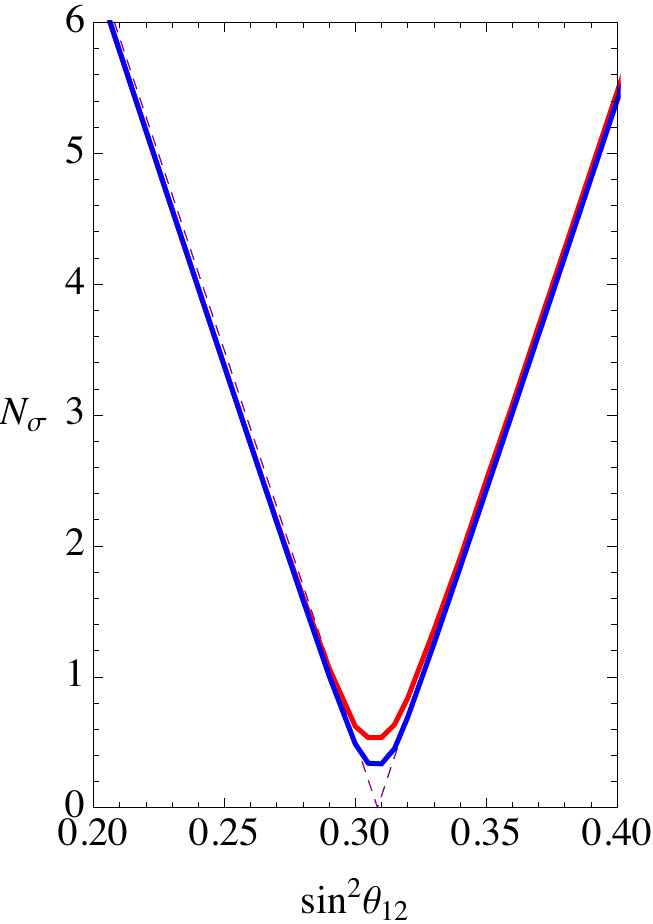}
 \vspace*{0.2cm}
 \fbox{\footnotesize Standard Ordering - Bimaximal} \\[0.5cm]
 \vspace*{-0.2cm} 
 \includegraphics[height=6.5cm]{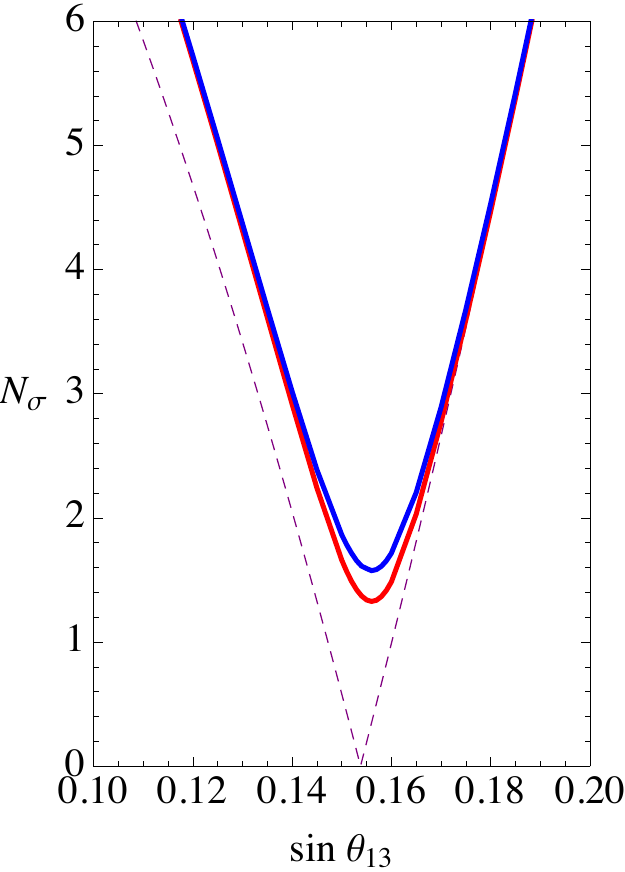} 
 \includegraphics[height=6.5cm]{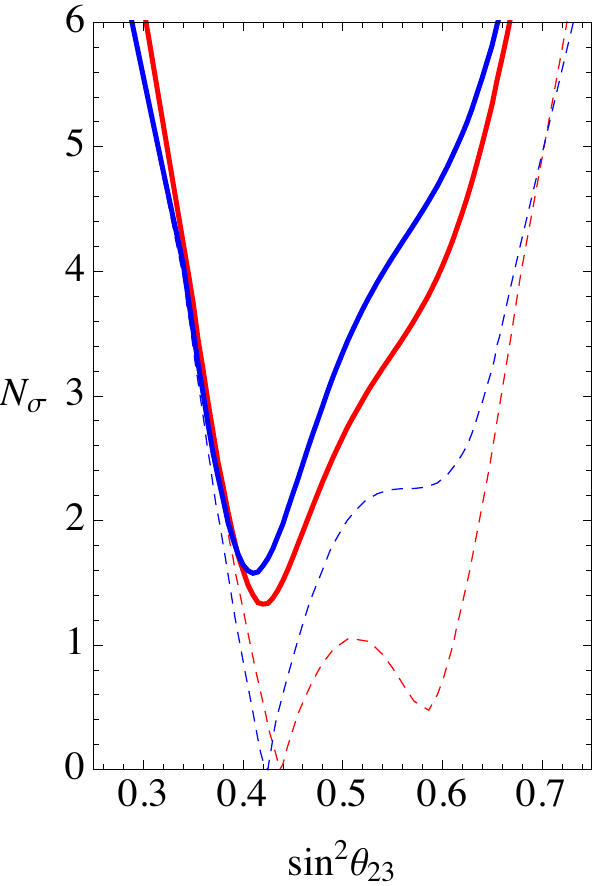}
 \includegraphics[height=6.5cm]{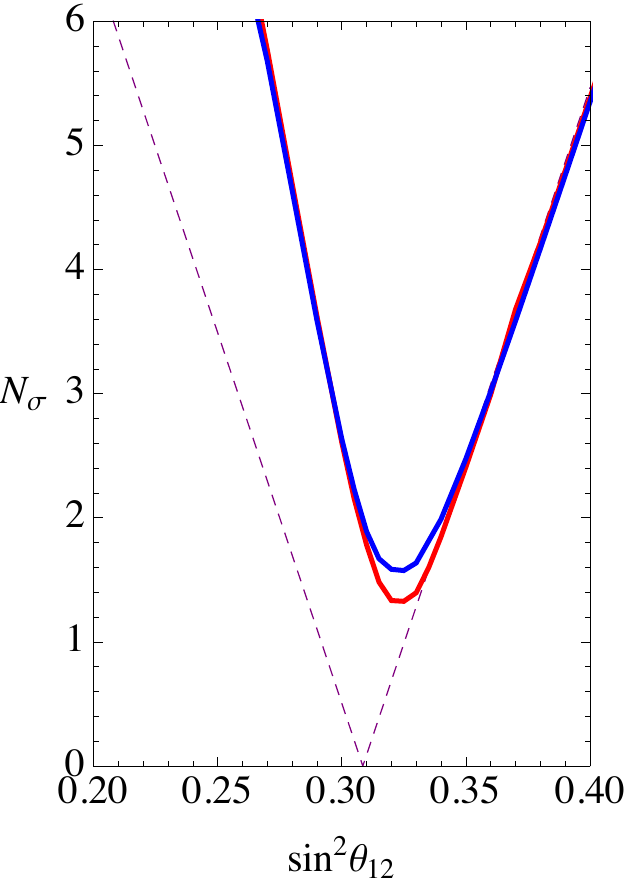}
 \caption{$N_\sigma$ as a function of each mixing angle for 
the TBM and BM models in the standard ordering setup.
The dashed lines represent the results of the global fit 
reported in \cite{Capozzi:2013csa} while the thick ones represent 
the results we obtain in our setup. Blue lines are 
for normal hierarchy while the red 
ones are for inverted hierarchy (we used purple when the two bounds are 
approximately identical). 
These bounds are obtained minimizing the value of $N_\sigma$ in 
the parameter space for fixed value of the showed mixing angle.}
\label{fig:Nsigma_bounds_StOrd_2013}
\end{figure}

\begin{figure}[p]
 \centering
 \vspace*{-1cm}
 \fbox{\footnotesize Standard Ordering - Tribimaximal} \\[0.5cm]
 \vspace*{-0.2cm} 
 \includegraphics[height=7cm]{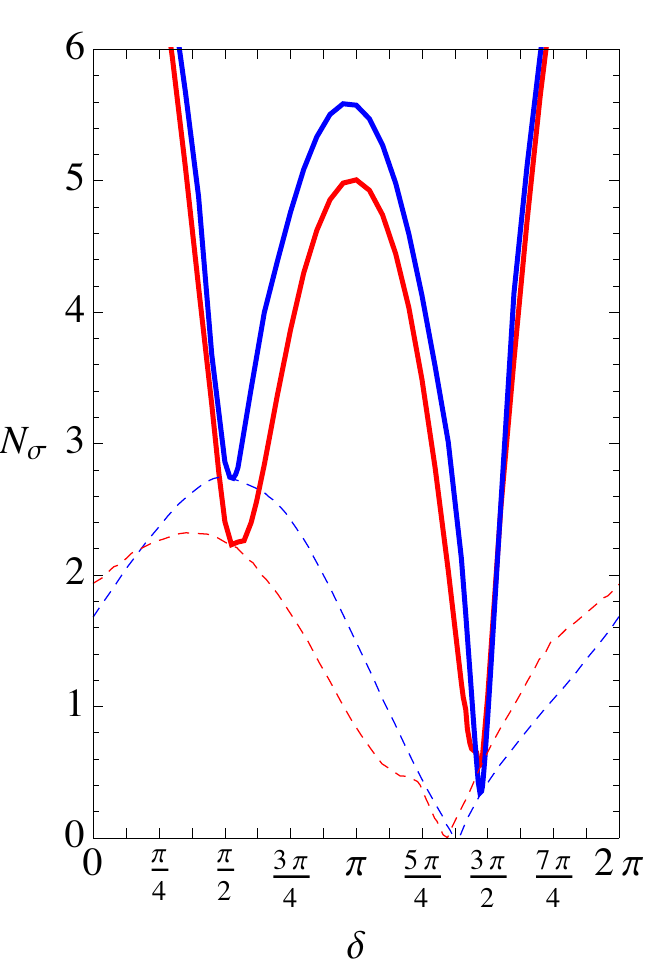} \hspace{1cm}
 \includegraphics[height=6.5cm]{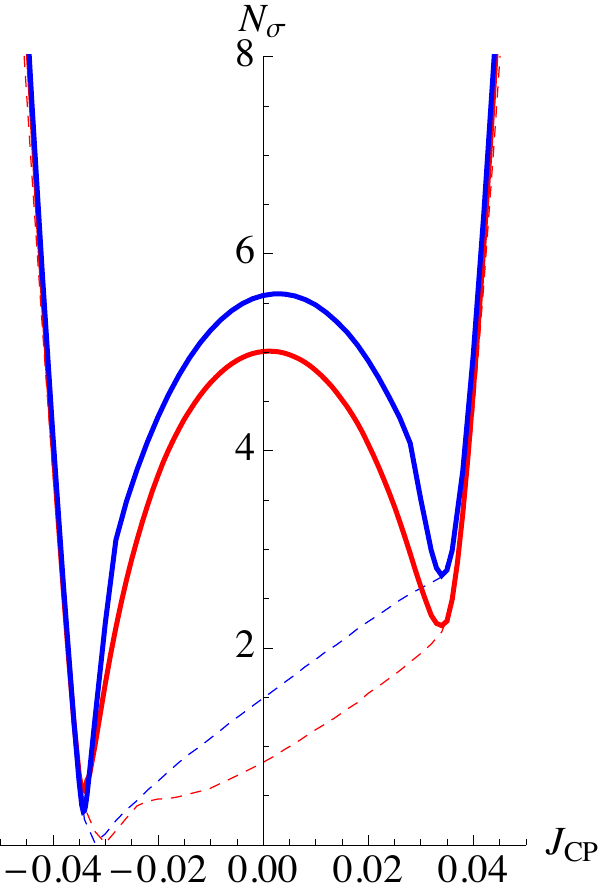}\\
 \vspace*{0.2cm}
 \fbox{\footnotesize Standard Ordering - Bimaximal} \\[0.5cm]
 \vspace*{-0.2cm} 
 \includegraphics[height=7cm]{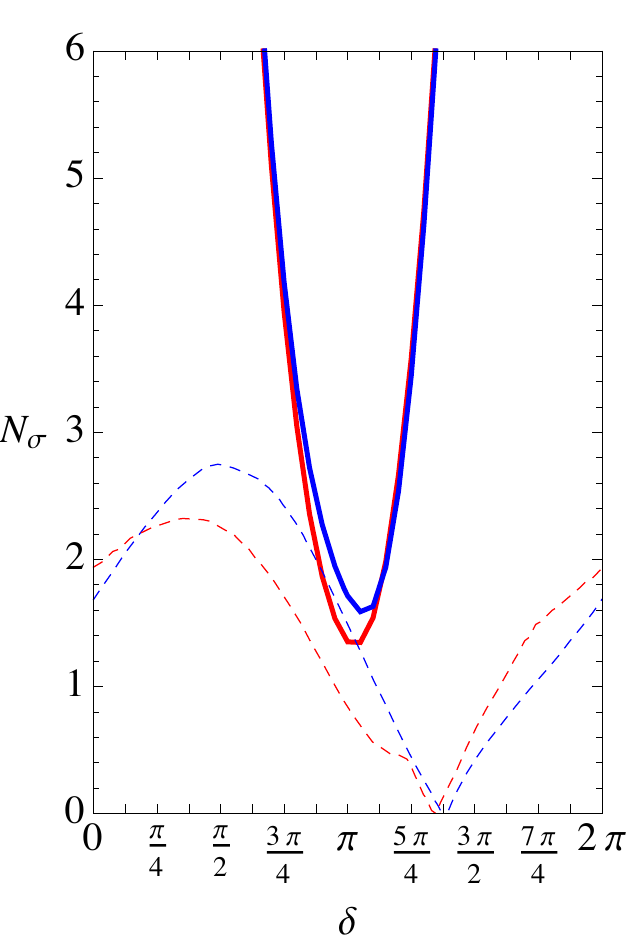} \hspace{1cm}
 \includegraphics[height=6.5cm]{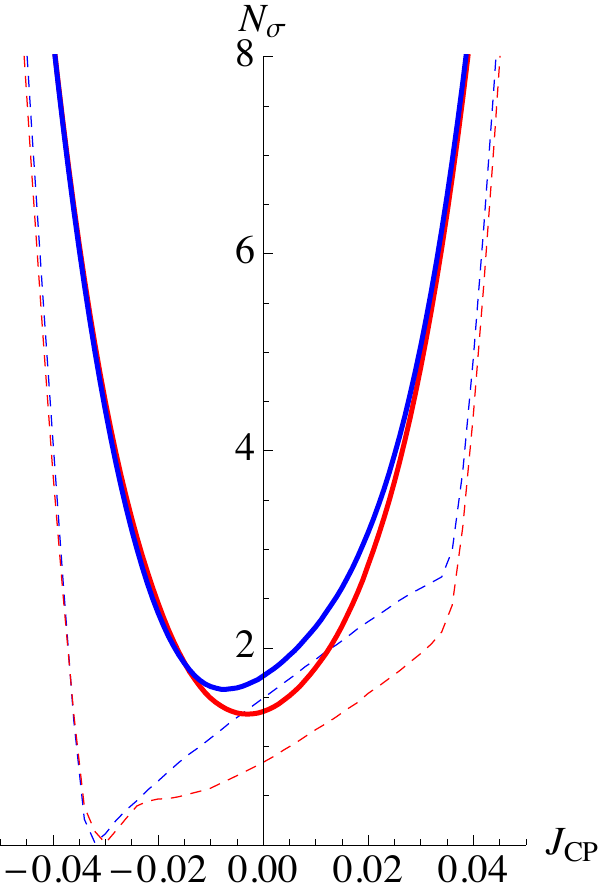}\\
\caption{\label{fig:bounds_Jcp_2013} 
\small
$N_\sigma$ as a function of $\delta$ and $J_{CP}$ for the TBM and BM models 
in the standard ordering setup.  The dashed lines represent 
the results of the global fit reported in \cite{Capozzi:2013csa} 
while the thick ones represent the results we obtain in our setup. 
Blue lines are for normal hierarchy while the red ones are 
for inverted hierarchy. These bounds are obtained minimizing 
the value of $N_\sigma$ in the parameter space for a 
fixed value of $\delta$ (left plots) or $J_{CP}$ (right plots).
}
\end{figure}

Using the new results from \cite{Capozzi:2013csa}
we find that the CP conserving values of $\delta =0,\pi$ ($J_{CP} = 0$)
are excluded in the tri-bimaximal mixing (TBM) case
at the level of $5\sigma$ both for the NH and IH spectra, 
as  Figs.~\ref{fig:results_delta_2013} 
and \ref{fig:bounds_Jcp_2013} show.
The second minimum at $\delta \sim \pi/2$ is disfavored with respect 
to the best-fit point at the $\sim 2.6\sigma$ ($2.2\sigma$) 
level in the NH (IH) case. The allowed $3\sigma$ ranges are 
reported in Table~\ref{table:bfAND3sigma_2013}.
In what concerns the atmospheric neutrino mixing angle, 
Fig.~\ref{fig:Nsigma_bounds_StOrd_2013} shows that, 
for TBM arising from the neutrino sector,
the allowed range of $\sin^2\theta_{23}$ in the model considered
follows  closely the results of the global fit in \cite{Capozzi:2013csa}: 
in the NH case we find a mild ($\sim 2\sigma$) preference for the first 
octant, while in the IH case the two octants are approximately degenerate.

We have found earlier that the value of $\delta$ predicted  
in the bimaximal mixing (BM) case using the best fit values of 
the neutrino mixing angles, is  $\delta \sim \pi$.
As a consequence, the BM case  is now mildly disfavored with 
respect to the TBM one:
\be
\sqrt{\chi^2_{min, BM} - \chi^2_{min, TBM}} \simeq 1.5 \; (1.2)\,,~~~
{\rm NH~(IH)}\,.
\ee
%
This is due to the new bound on $\delta$ obtained in 
\cite{Capozzi:2013csa}, see Fig.~\ref{fig:bounds_Jcp_2013}.
For the atmospheric neutrino mixing angle $\theta_{23}$
we find in this case a preference for 
the first octant at the $\sim 3 \sigma$ level, 
see Table~\ref{table:bfAND3sigma_2013} 
and Figs.~\ref{fig:results_delta_2013}, 
\ref{fig:Nsigma_bounds_StOrd_2013} .

%
%
\begin{table}[t]
\centering
\begin{tabular}{l l | c | c }
\toprule
		& 	& Best fit & 3$\sigma$ range  \\ \hline
		& $J_{CP}$ (NH)  & $-0.034$ & $-0.038 \div -0.028 \oplus 0.032 \div 0.036 $ \\
		& $J_{CP}$ (IH)  & $-0.034$ & $-0.039 \div -0.024 \oplus 0.027 \div 0.037 $ \\  
		& $\delta$ (NH)  & $4.63$ & $1.53 \div 1.80 \oplus 4.24 \div 4.92 $ \\  
	TBM	& $\delta$ (IH)  & $4.62$ & $1.45 \div 2.10 \oplus 4.03 \div 4.94 $ \\  
		& $\sin \theta_{13}$ & $0.15$ &  $0.13 \div 0.17$  \\ 
		& $\sin^2 \theta_{23}$ (NH) & $0.43$ &  $0.36 \div 0.64$  \\
		& $\sin^2 \theta_{23}$ (IH) & $0.44$ &  $0.36 \div 0.66$  \\ 
		& $\sin^2 \theta_{12}$ & $0.31$ &  $0.26 \div 0.36$  \\ \hline
		
		& $J_{CP} (NH)$ & $-0.008$ & $-0.026 \div 0.022$ \\ 
		& $J_{CP} (IH)$ & $-0.003$ & $-0.025 \div 0.023$ \\ 
		& $\delta$ (NH)  & $3.35$ & $2.50 \div 3.92 $ \\  
	BM	& $\delta$ (IH)  & $3.22$ & $2.47 \div 3.88 $ \\  
		& $\sin \theta_{13}$ & $0.16$ &  $0.14 \div 0.17$  \\ 
		& $\sin^2 \theta_{23}$ (NH) & $0.41$ &  $0.35 \div 0.50$  \\
		& $\sin^2 \theta_{23}$ (IH) & $0.42$ &  $0.36 \div 0.55$  \\ 
		& $\sin^2 \theta_{12}$ & $0.32$ &  $0.29 \div 0.36$  \\
\bottomrule
\end{tabular}
\caption{\label{table:bfAND3sigma_2013} Best fit and 3$\sigma$ ranges 
(found fixing $\sqrt{\chi^2 - \chi^2_{min}} = 3$) in the standard 
ordering setup using the data from \cite{Capozzi:2013csa}. 
When not explicitly indicated 
otherwise, the result applies both for normal hierarchy and inverted hierarchy 
of neutrino masses.}
\end{table}

%

\newpage
\label{Addendum_out}

\end{document}